\documentclass[useAMS,usenatbib]{mn2e}
\usepackage{graphicx}
\def\beg{\begin{eqnarray}}
\def\ende{\end{eqnarray}}
\def\gsim{\lower.4ex\hbox{$\;\buildrel >\over{\scriptstyle\sim}\;$}} 
\def\lsim{\lower.4ex\hbox{$\;\buildrel <\over{\scriptstyle\sim}\;$}}

\renewcommand{\vec}[1]{\mbox{\boldmath $#1$}}
\def\curl{{\rm curl}} 

\title[Radiation zone dynamo]
{Helicity  and dynamo action  in magnetized stellar radiation zones}
\author[G. R\"udiger, L. L. Kitchatinov and  D. Elstner]
{
G. R\"udiger$^1$\thanks{E-mail: GRuediger@aip.de},
L. L. Kitchatinov$^{1,2,3}$,  
D. Elstner$^1$ 
\\
$^1$Leibniz-Institut f\"ur Astrophysik Potsdam, An der Sternwarte 16, 14482 Potsdam, Germany\\
$^2$Institute for Solar-Terrestrial Physics, PO Box 291, Irkutsk 664033, Russia\\
$^3$Pulkovo Astronomical Observatory, St. Petersburg, 196140, Russia
}
\begin{document}

\date{Accepted . Received ; in original form }

\pagerange{\pageref{firstpage}--\pageref{lastpage}} \pubyear{2009}

\maketitle

\label{firstpage}

\begin{abstract}
Helicity and $\alpha$ effect  driven by the nonaxisymmetric  Tayler instability of toroidal magnetic fields in  stellar radiation zones are computed. In the linear approximation a purely toroidal field always excites  pairs of modes  with identical   growth rates but with opposite helicity so that the net helicity  vanishes. If the magnetic background field has a helical structure by an extra  (weak) poloidal component then one of the modes dominates producing a net kinetic helicity anticorrelated to the current helicity of the background field.\\
The mean electromotive force is computed with the  result that the  $\alpha$ effect  by   the most rapidly growing  mode has the same sign as the current helicity of the background field. The  $\alpha$ effect is found as  too small to drive an $\alpha^2$ dynamo but  the excitation conditions for an $\alpha\Omega$ dynamo can     be fulfilled  for  weak poloidal fields. Moreover, if  the dynamo produces its own $\alpha$ effect by the magnetic instability then   problems with its  sign do not arise. For all cases,  however,  the $\alpha$ effect    shows an extremely  strong concentration to the poles so that a possible $\alpha\Omega$ dynamo might only work at the polar regions. Hence,   the  results of our linear theory lead to a new topological problem for the existence of  large-scale dynamos in  stellar radiation zones on the basis of the current-driven instability of toroidal fields.
\end{abstract}

\section{Introduction}
Hydromagnetic dynamos can be understood   as magnetic instabilities driven by  a special flow pattern in fluid conductors.  There are, however, strong restrictions on the characteristics  of such flows (see Dudley \& James 1989) as well as on the geometry of the resulting magnetic fields  \citep{C33}. The restrictions even exclude any dynamo activity  for a number of flows.  We mention as  an example that differential rotation alone can never maintain a dynamo (Elsasser 1946). 

An open question is whether  magnetic instabilities are able to  excite   a sufficiently complicated motion that together with a (given)  background flow can generate magnetic fields. \citet{TP92} suggested that nonuniformly rotating disks can produce a dynamo when magnetorotational (MRI) and magnetic buoyancy instabilities are active. Later on, numerical simulations of \citet{Bea95} and \citet{Hea96} have shown that MRI alone may be sufficient for the accretion disk dynamo. It remains, however, to check (at least for the case of low magnetic Prandtl number) whether the MRI-dynamo has physical or numerical origin \citep{FP07,Fea07}.  

Another possibility was discussed by \citet{S02} who suggested that differential rotation and magnetic kink-type instability \citep{T73} can jointly drive a dynamo in stellar radiation zones. The dynamo if real would be very important for the angular momentum transport in stars and their secular evolution. It taps energy from differential rotation thus reducing the rotational shear. Radial displacements converting toroidal magnetic field into poloidal field are necessary for the dynamo. The dynamo, therefore, unavoidably mixes chemical species in stellar interiors that may have observable consequences for stellar evolution. 

 Such a dynamo, however,  has not yet been  demonstrated to exist. The doubts especially concern  the kink-type instability that in contrast to MRI exists also without differential rotation. The Tayler instability develops in expense of magnetic energy. Estimations of dynamo parameters are thus necessary to assess  the dynamo-effectiveness of this magnetic instability. 

The basic role in turbulent dynamos plays  the ability of correlated magnetic (${\vec b}$) fluctuations and velocity (${\vec u}$) fluctuations to produce a mean electromotive force along the background magnetic field ${\vec B}$ and also along the electric current $\vec J$, i.e.
\begin{equation}
  \langle {\vec u}\times{\vec b}\rangle = \alpha{\vec B}- \mu_0 \eta_{\rm T}{\vec J} .
  \label{1}
\end{equation}
We estimate   the $\alpha$ effect by Tayler instability in the present paper. We do also find   indications for the appearance of the turbulent diffusivity $\eta_{\rm T}$ in the calculations but  we do not follow them here in detail. For purely toroidal fields we did {\em not} find indication for the existence of the 
 term ${\vec \Omega}\times \vec{J}$ which can appear in the expression (\ref{1}) in form  of a rotationally induced anisotropy of the diffusivity tensor.

The fluctuating fields for the most rapidly growing eigenmodes and the azimuthal averaging are applied in the LHS of Eq.\,(\ref{1}) to estimate the $\alpha$ effect and its relation to the kinetic and magnetic helicity ${\cal H}^{\rm kin}$ and ${\cal H}^{\rm curr}$. Our linear stability computations do not allow the evaluation of the $\alpha$ effect amplitude but its latitudinal profile  and its ratio to the product of rms values of $\vec{u}$ and $\vec{b}$ (i.e. the  correlation coefficient) can be found. As the differential rotation is necessary for dynamo, we estimate also the influence of differential rotation  on Tayler instability. Next, a dynamo model with the parameters estimated for the magnetic instability is designed to find the global modes of the instability-driven dynamo.
\section{Instability and helicity}
The model and the  stability analysis of this paper are very close to that of \citet{KR08} and will be discussed here only briefly. 

The basic  component of the magnetic field inside a  star is normally assumed to be the toroidal one. This toroidal field can be produced by differential rotation from even a small poloidal field. The background toroidal field of our model consists of two latitudinal belts of opposite polarities, i.e.
\begin{equation}
   {\vec B} =  r\sin\theta\cos\theta\sqrt{\mu_0\rho}\
   \Omega_\mathrm{A} {\vec e}_\phi 
   \label{2}
\end{equation} 
(see Spruit 1999) with   $\Omega_\mathrm{A}$ as the Alfv\'en  frequency of the toroidal field. Spherical coordinates are used with the axis of rotation as the polar axis and  ${\vec e}_\phi$ as the azimuthal unit vector. The latitudinal profile of (\ref{2}) peaks in mid-latitudes at $\theta=45^\circ$ and $\theta=135^\circ$.

The background flow is simply
\begin{equation}
  {\vec U} =  r\sin\theta \ \Omega \ {\vec e}_\phi
  \label{3}
\end{equation}
with  $\Omega$  as the equatorial rotation rate. 

The $\Omega_\mathrm{A}$ and $\Omega$ are  radius-dependent but this dependence is not of basic importance for the stability analysis. The reason is that  the  stratification of the radiative core is stable with positive  
\begin{equation}
   N^2 = \frac{g}{C_\mathrm{p}}\frac{\partial S}{\partial r}  ,
   \label{4}
\end{equation}
where $S$ is the entropy and $C_\mathrm{p}$ is the specific heat at constant pressure. 

The buoyancy frequency  $N$ is large compared to $\Omega$ ($N/\Omega \simeq 400$ in the upper radiative core of the Sun). Then the radial scale of unstable disturbances is short and the dependence of the disturbances on radius can be treated in a local approximation, i.e. in the form of $\mathrm{exp}(\mathrm{i}kr)$. The parameter controlling the stratification influence on the instability is 
\begin{equation}
   \hat\lambda = \frac{N}{\Omega k r} , 
   \label{5}
\end{equation}
and the most unstable disturbances have $\hat\lambda < 1$ \citep{KR08}. This means that the radial scale of the disturbances, $\lambda = \pi/k$, is short compared to the radial scale $\delta R$ of toroidal field or angular velocity variations. In the solar tachocline where $\Omega$ strongly varies  in radius, the scale ratio is smaller than unity, $\lambda /\delta R \simeq 0.2$ \citep{KR09}. For such  small scale ratio the radial derivatives  in the linear stability equations are absorbed by the disturbances so that the  local approximation in the  radial coordinate  can be applied. Note that the unstable modes remain global in horizontal dimensions. 
\begin{figure}
  \includegraphics[width=4.1cm,height=4.5cm]{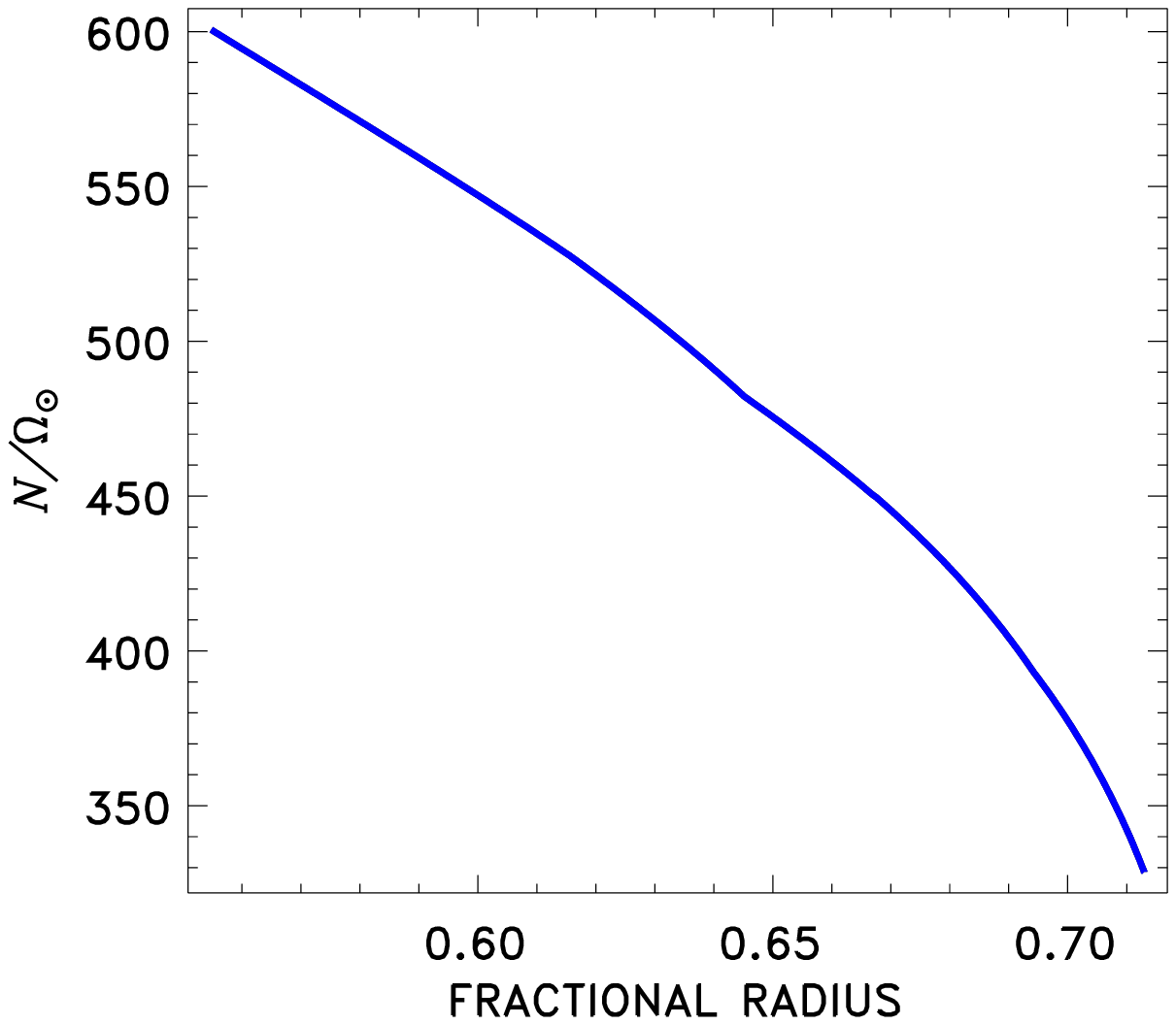}
  \hspace{0.1truecm}
  \includegraphics[width=4.1cm,height=4.5cm]{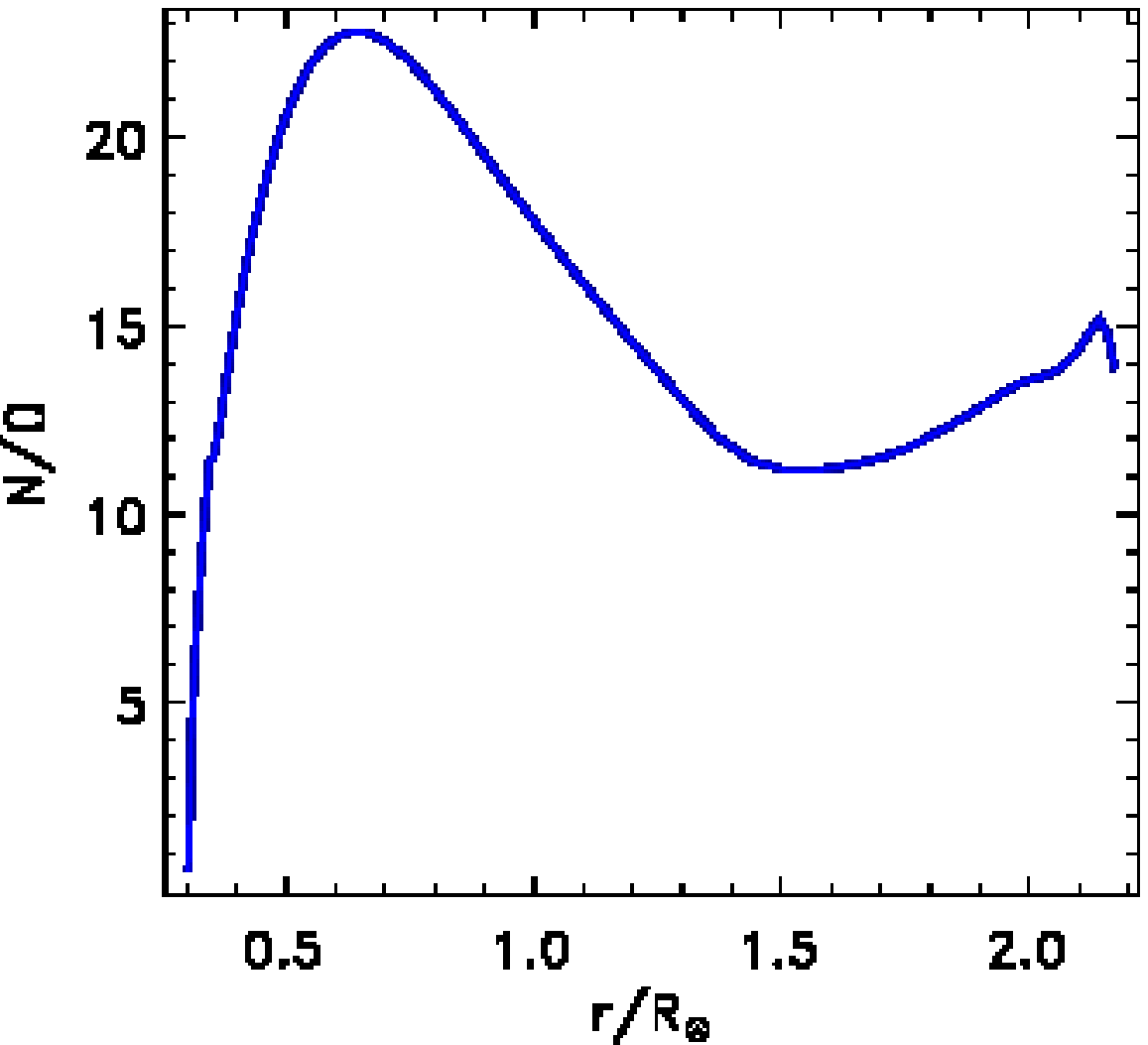}
  \caption{The ratio  $N/\Omega_0$  in the radiation 
  zones of the Sun ({\em left}) and  of  a 100~Myr old $3M_\odot$ star 
  rotating with  10 days period ({\em right}). The model of the massive star is  computed with the EZ code of \citet{P04} for  $X=0.7,\ Z=0.02$.
  }
  \label{f5}
\end{figure}

\subsection{Instability of toroidal fields}

The equation system of the linear stability analysis include 
\begin{eqnarray}
 \frac{\partial{\vec u}}{\partial t}
 + \left({\vec U}\cdot\nabla\right){\vec u}
 + \left({\vec u}\cdot\nabla\right){\vec U}
 + \frac{1}{{\mu_0\rho}}\left(\nabla\left({\vec B}\cdot{\vec b}\right)\right.&&
 \nonumber \\
  - \left. \left({\vec B}\cdot\nabla\right){\vec b}
 - \left({\vec b}\cdot\nabla\right){\vec B}\right)
 =-\left(\frac{1}{\rho}\nabla P\right)' + \nu\Delta{\vec u} &&
  \label{6}
\end{eqnarray}
as the momentum equation for the velocity fluctuations and the   equations
 \begin{equation}
   \frac{\partial{\vec b}}{\partial t} =
   \nabla\times\left( {\vec U}\times{\vec b}
   + {\vec u}\times{\vec B} - \eta\nabla\times{\vec b}\right) ,
   \label{7}
\end{equation}
for the induction of the magnetic fluctuations and for the disturbances ($s$) of the entropy
\begin{equation}
  \frac{\partial s}{\partial t} + {\vec U}\cdot \nabla s +
  {\vec u}\cdot\nabla S = \frac{C_\mathrm{p}\chi}{T}\Delta T'. 
  \label{8}
\end{equation}
 The equations (\ref{6})--(\ref{8}) were reformulated in terms of scalar potentials for toroidal and poloidal parts of the magnetic and velocity fields  to reduce the number of equations, i.e.
\begin{eqnarray}
  \vec{u} =-\curl((rW)\vec{r}) -\curl\ \curl((rV)\vec{r})&&
 \nonumber \\
  \vec{b} = -\curl ((rB)\vec{r}) -\curl\ \curl((rA)\vec{r})\ \ &&
  \label{9b}
\end{eqnarray}
(Chandrasekhar 1961). The resulting system of five eigenvalue equations can be found elsewhere \citep{KR08} and it will here only be  extended  to the application of   helical background fields. Again the  eigenvalue problem is solved numerically.   

The equations include finite diffusion in opposition to the otherwise similar equations of Cally (2003). The thermal diffusion is especially important because of its destabilizing effect. The Tayler instability requires radial displacements. It does not exist in 2D case of strictly horizontal motion. The radial displacements in radiation zones are opposed by buoyancy. The thermal diffusion smooths out the entropy disturbances to reduce the effect of stable stratification. This largely increases the growth rates for the instability of not too strong, $\Omega_\mathrm{A} < \Omega$, fields \citep{RK09}. The diffusivities enter the normalized equation via parameters
\begin{equation}
  \epsilon_\chi = \frac{\chi N^2}{\Omega_0^3 r^2},\ \ \ \ \ \ \
  \epsilon_\eta = \frac{\eta N^2}{\Omega_0^3 r^2},\ \ \ \ \ \ \
  \epsilon_\nu = \frac{\nu N^2}{\Omega_0^3 r^2}.
  \label{9}
\end{equation}
In our computations we used the values  $\epsilon_\chi = 10^{-4}$, $\epsilon_\eta = 4\times 10^{-8}$, and $\epsilon_\nu = 2\times 10^{-10}$ characteristic for the upper part of the solar radiation zone. As it must be the magnetic Prandtl number ($5\cdot 10^{-3}$) exceeds the ordinary Prandtl number 
($2\cdot 10^{-6}$).

The stability problem allows two types of equatorial symmetry of unstable eigenmodes. We use the notations Sm (symmetric mode with azimuthal wave number $m$) and A$m$ (antisymmetric mode) for these types of symmetry. S$m$ modes have vector field $\vec{b}$ which is mirror-symmetric about equatorial plane (symmetric ${b_r}, {b_\phi}$  and antisymmetric ${b_\theta}$) and mirror-antisymmetric flow $\vec{u}$ (symmetric ${u_\theta}$ and antisymmetric ${u_r}, {u_\phi}$). Am modes have antisymmetric $\vec{b}$ and symmetric $\vec{u}$. Instability can only be  found for nonaxisymmetric disturbances with $m=\pm 1$  in agreement with the stability criteria of \citet{Gea81} when applied to the toroidal field model of Eq.~(\ref{2}).  

\begin{figure}
  \includegraphics[width=7cm,height=7cm]{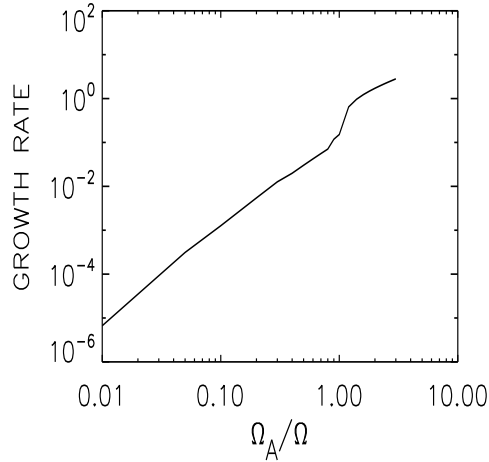}
   \caption{The normalized growth rate $\hat \gamma=\Im{\omega}/\Omega$ of S modes   as function of the amplitude of the  toroidal background  field. Rigid rotation,   $m=\pm 1$. Note the  slowness of the modes for weak fields  with $\Omega_{\rm A}<\Omega$ and the jump at $\Omega_{\rm A}\simeq\Omega$. The wave numbers are always   optimized with respect to the maximum 
growth rates. ${\rm Pm}=5\cdot 10^{-3}$, ${\rm Pr}=2\cdot 10^{-6}$.}
  \label{f1}
\end{figure}

The perturbations are considered as Fourier modes in time, in azimuth and radius in the form $\exp({\rm i}(kr+m\phi -\omega t))$. Only the highest-order terms in $kr$ are used so that in radial direction the theory is a local one (short-wave approximation). The wave number $k$ enters the equations in the normalized form 
\begin{equation}
  \hat \lambda= \frac{N}{\Omega_0 kr}
  \label{lambda}
\end{equation}
as a ratio of two large numbers. If only toroidal fields are considered with $\Omega_{\rm A}/\Omega=0.1$ Kitchatinov \& R\"udiger (2008) found maximal growth  rates  of order 10$^{-4}$ (normalized with the rotation rate) at radial scales of $\hat\lambda\simeq 0.1$ for S1 modes.

As described in KR08, the eigenvalue $\omega$ possesses a positive imaginary part  for an instability (the growth rate is $\gamma=\Im(\omega)$). The equations have only be solved for the nonaxisymmetric (`kink') modes with $m=\pm 1$. One can easily show that 
the equation system possesses  a symmetry with respect to the change of sign of the azimuthal wave number $m$. The equations are invariant under the transformation
\begin{eqnarray}
  &&\left( m,\hat{\omega},W,V,B,A,S\right) \rightarrow 
  \nonumber \\
  && \rightarrow 
  \left( -m, -\hat{\omega}^*, -W^*, V^*, -B^*,A^*,-S^*\right) ,
  \label{trans}
\end{eqnarray}
where the asterisks  mean the complex conjugate. If an eigenmode exists for $m=1$ with a certain growth rate $\gamma$ then the same $\gamma$ holds for the eigenmode with $m=-1$. We shall demonstrate the importance of this finding for the generation of helicity and $\alpha$ effect.

Figure~\ref{f1} shows the  growth rates for S$\pm 1$ modes in dependence of field strength. The results for A$\pm 1$ modes are very similar. For weak fields, $0.01 < \Omega_\mathrm{A}/\Omega < 1$ the growth rates are closely reproduced by the parabolic law $\gamma \simeq 0.1 \Omega^2_\mathrm{A}/\Omega$. For strong fields, $\Omega_\mathrm{A} > \Omega$, they are proportional to the field strength,
$\gamma \simeq \Omega_\mathrm{A}$ and do not depend on the rotation rate.

\subsection{Helicity production}
\begin{figure}
 \vbox{
\includegraphics[width=7cm,height=6cm]{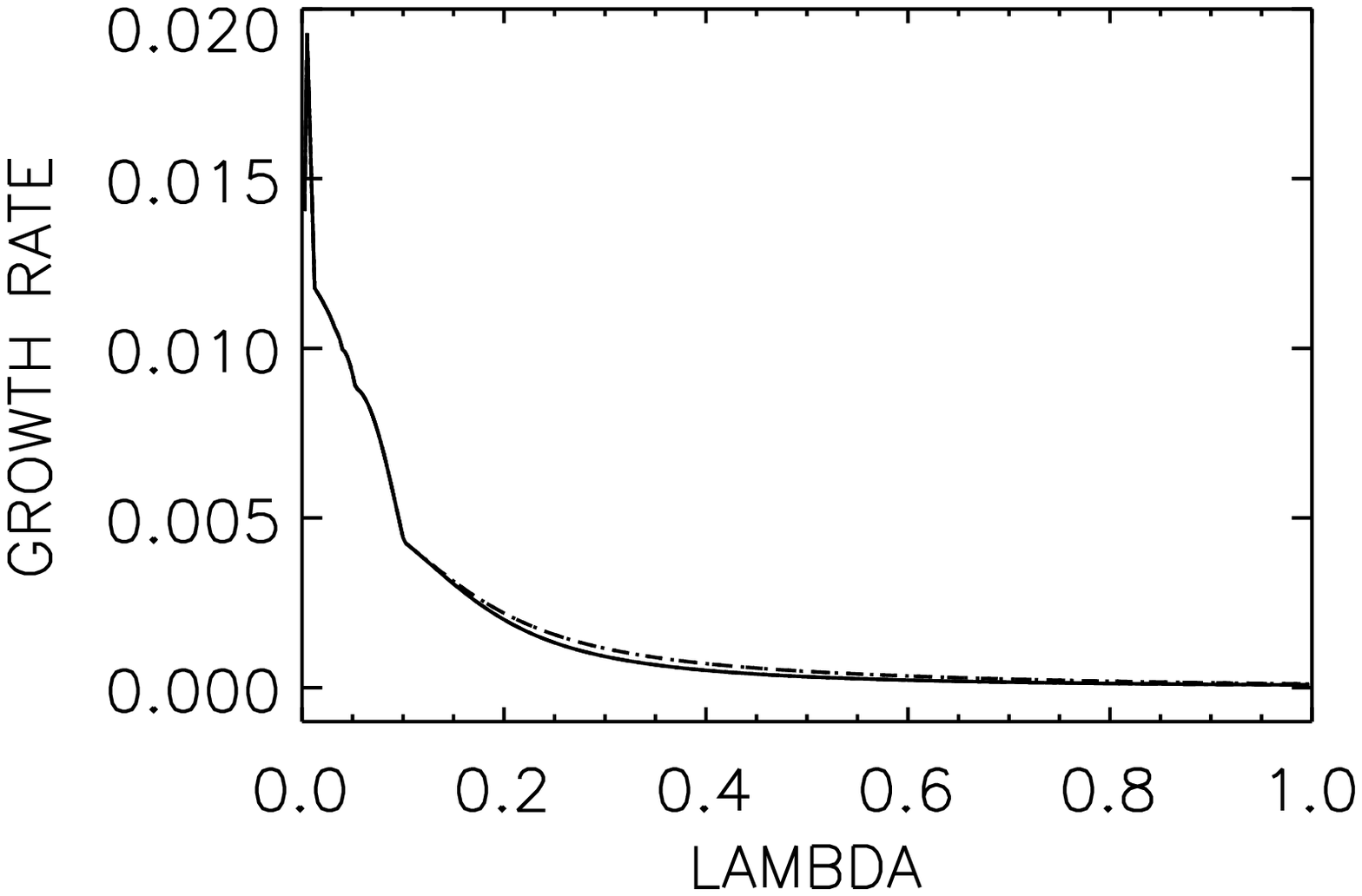}
\includegraphics[width=7cm,height=6cm]{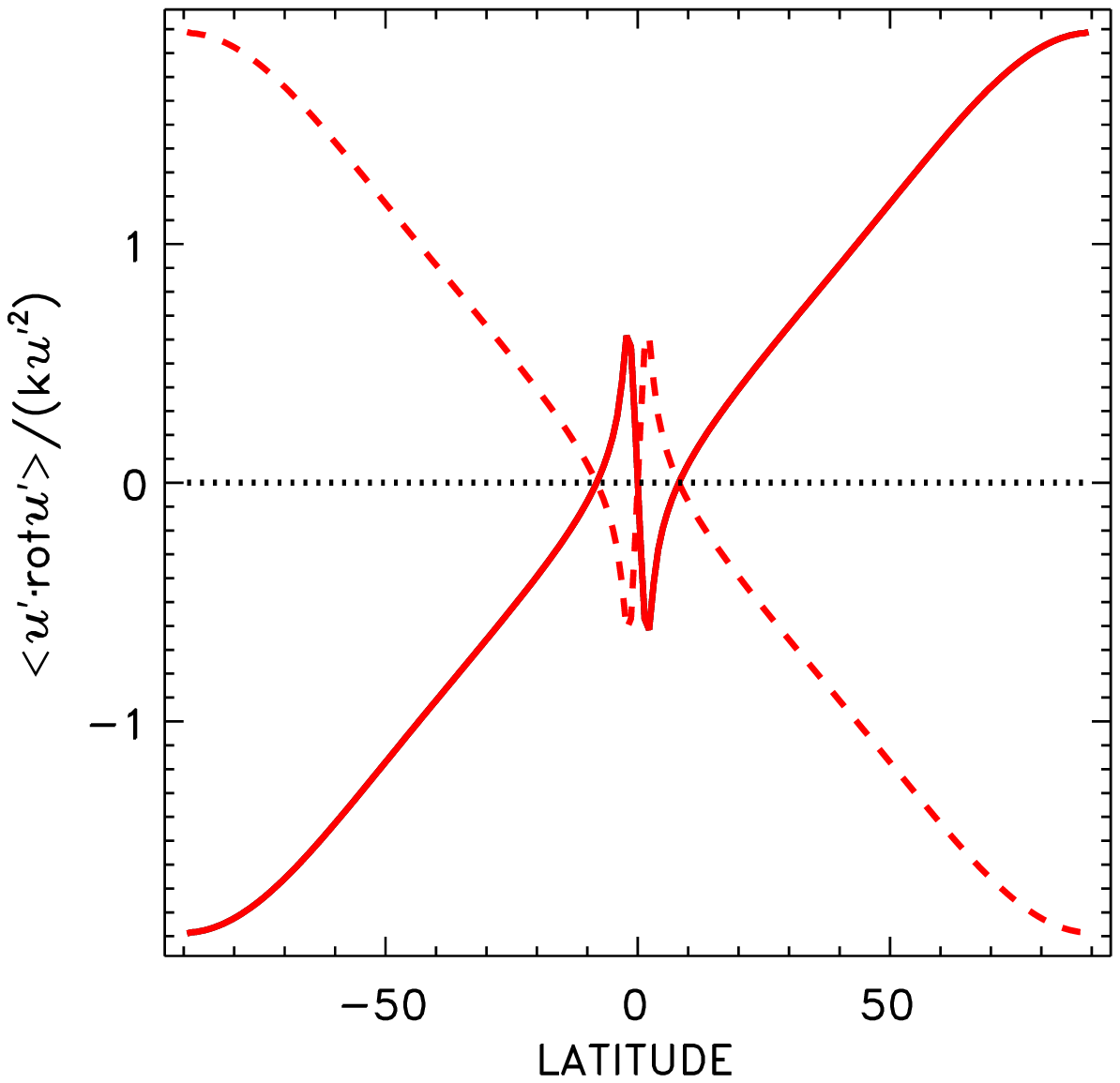}
}
 \caption{Purely toroidal fields for rigid rotation. Top: The growth rate $\hat\gamma$ for $\hat\Omega_{\rm A}=0.3$ vs. the radial scale $\hat\lambda$ of the perturbations $A\pm1$ and $S\pm1$. Bottom: The normalized small-scale helicity (\ref{Hkin}). Solid line: $m=1$, dashed line $m=-1$. The net helicity vanishes if both modes are simultaneously excited. $B_r=0, \ \hat\lambda = 0.1$.}
  \label{f10}
\end{figure}

 All averaging procedures in the present paper are  realized by integration over the azimuth coordinate.
Consider the formation of the kinetic helicity
\beg
\cal{H}^{\rm kin} = \langle \vec{u}\cdot \curl\ {\vec u}\rangle,
\label{Hkin}
\ende
(only the real parts of both factors) whose definition  depends on the handedness of the used coordinate system. We shall prefer the righthand system.   Let the expressions
\beg
W= (w_{\rm R} + {\rm i}  w_{\rm I}){\rm e}^{{\rm i}\phi},\ \ \ \ \ \  V= (v_{\rm R} + {\rm i}  v_{\rm I}){\rm e}^{{\rm i}\phi},  
\label{wv}
\ende
represent the   potential functions $W$ and $V$. Then the real part of (\ref{Hkin}) after the integration over the azimuth results to 
\begin{eqnarray}
\lefteqn{ {\cal H}^{{\rm kin}} = \frac{1}{\sin^2\theta}  \bigg[\bigg(m w_{\rm I} + \sin\theta \frac{\partial v_{\rm I}}{\partial\theta}\bigg)\bigg(m v_{\rm I} + \sin\theta \frac{\partial w_{\rm I}}{\partial\theta}\bigg)+}\nonumber\\
&& + \bigg(m w_{\rm R} + \sin\theta \frac{\partial v_{\rm R}}{\partial\theta}\bigg)\bigg(m v_{\rm R} + \sin\theta \frac{\partial w_{\rm R}}{\partial\theta}\bigg)\bigg].
\label{Hkin1}
\end{eqnarray}
It is easy to show with  the transformation rules (\ref{trans}) that  this expression  for the modes with negative $m$  has the opposite sign as for the modes with positive  $m$.
Hence,
\beg
 {\cal{H}}^{\rm kin}(m=-1) = - {\cal{H}}^{\rm kin}(m=1)
\label{Hkin2}
\ende
(see Fig.~\ref{f10}, bottom). It means that for every unstable mode with finite helicity there is another unstable mode with the same growth and drift rates but with opposite  helicity so that the resulting net  helicity should vanish. Obviously, if all modes are excited the instability of a purely toroidal axisymmetric field can not produce finite values of the kinetic helicity. The same  argument leads to the same conclusion for the current helicity 
\beg
\cal{H}^{\rm curr} = \langle \vec{b}\cdot \curl\ {\vec b}\rangle.
\label{Hcurr}
\ende
There is also a more straightforward argument with the same result. The mode $m=-1$ is identical to the mode $m=1$ but considered in a lefthand coordinate system. The sign of the helicity is equal in both lefthand systems and righthand systems. Hence, the mode $m=-1$ which gives the same helicity in the lefthand system as the mode $m=1$ in the righthand system yields a negative helicity in the righthand system if  the mode $m=1$ yields a positive helicity in the righthand system. The net helicity in both the righthand system and the lefthand  system, therefore, vanishes .
Again the  same is true for the current helicity (\ref{Hcurr}).

\section{Influence of poloidal  fields}
We find (see, however, Cally 2003) that from  symmetry reasons unstable purely toroidal fields do not produce a net helicity.
An additional  poloidal component of the background field, however,   breaks the symmetry as then,  for example, the two modes possess two different growth rates  so that a certain sign of helicity will be  preferred (Gellert, R\"udiger \& Hollerbach 2011). 

In the  short-wave approximation only the radial component of the poloidal field is important. We write
\begin{equation}
B_r= r \sqrt{\mu_0\rho}\ \Omega_{\rm A,p}(\mu)
\label{BR}
\end{equation}
with $\mu=\cos\theta$. Using the notation  by KR08 with the operator
 \begin{equation}
  {\hat{\cal L}} = \frac{1}{\sin\theta}\frac{\partial}{\partial\theta}
  \sin\theta\frac{\partial}{\partial\theta} +
  \frac{1}{\sin^2\theta}\frac{\partial^2}{\partial\phi^2},
\label{LL}
\end{equation}
the equation for the azimuthal flow reads
\begin{eqnarray}
  &&\hat\omega\left(\hat{\cal L}W\right) =
  - \mathrm{i}\frac{\epsilon_\nu}{\hat\lambda^2}\left(\hat{\cal L}W\right)
  + m\hat\Omega\left(\hat{\cal L}W\right)
  - m\hat\Omega_\mathrm{A}\left(\hat{\cal L}B\right)
  \nonumber \\
  &&- mW\frac{\partial^2}{\partial\mu^2}
  \left(\left(1-\mu^2\right)\hat\Omega\right)
  + mB\frac{\partial^2}{\partial\mu^2}
  \left(\left(1-\mu^2\right)\hat\Omega_\mathrm{A}\right)
  \nonumber \\
  &&+ \left(\hat{\cal L}V\right)\frac{\partial}{\partial\mu}
  \left(\left(1-\mu^2\right)\hat\Omega\right)
  - \left(\hat{\cal L}A\right)\frac{\partial}{\partial\mu}
  \left(\left(1-\mu^2\right)\hat\Omega_\mathrm{A}\right)
  \nonumber \\
  &&+ \left(\frac{\partial}{\partial\mu}\left(\left(1-\mu^2\right)^2
  \frac{\partial\hat\Omega}{\partial\mu}\right) -
  2\left(1-\mu^2\right)\hat\Omega\right)\frac{\partial V}{\partial\mu}
  \nonumber \\
  &&- \left(\frac{\partial}{\partial\mu}\left(\left(1-\mu^2\right)^2
  \frac{\partial\hat\Omega_\mathrm{A}}{\partial\mu}\right) -
  2\left(1-\mu^2\right)\hat\Omega_\mathrm{A}\right)
  \frac{\partial A}{\partial\mu}
  \nonumber \\
  &&-\frac{1}{\hat\lambda}\!\!\left(\!\hat\Omega_\mathrm{A,p} \left(\hat{\cal L}B\right) 
  \!+\! (1 \!-\! \mu^2) \frac{\partial\hat\Omega_\mathrm{A,p}}{\partial\mu}
  \frac{\partial B}{\partial\mu} \!-\! 
  m\frac{\partial\hat\Omega_\mathrm{A,p}}{\partial\mu} A\!\!\right).
  \label{tflow}
\end{eqnarray}
The equation includes the poloidal background field via its last line. The parameter measuring the effect of the poloidal field is 
\begin{equation}
  \hat{\Omega}_\mathrm{A,p} =  \frac{N}{\Omega} \frac{\Omega_{\rm A,p}}{\Omega}=\frac{N B_r(\mu )}{\Omega_0^2 r \sqrt{\mu_0\rho}}. 
  \label{om_p}
\end{equation}
Equation (\ref{om_p}) shows that the characteristic strength of the field that can influence the Tayler instability ($\hat{\Omega}_\mathrm{A,p} \sim 1$) is $\Omega_0/N$ times the toroidal field amplitude. This factor is of the order $10^{-(2\dots 3)}$.

For the latitudinal profile of the radial field component the simplest choice, i.e. $B_r(\mu)\sim \mu$, is used. Hence, both the background field components $B_r$ and $B_\phi$   are antisymmetric with respect to the equator. The large-scale current helicity ${\vec B}\cdot \curl {\vec B}$, therefore, is also antisymmetric with respect to the equator. For positive amplitudes $\Omega_{\rm A}$ and $\Omega_{\rm A,p}$  it is {\em positive} at the northern  hemisphere  and negative at the southern hemisphere (it runs with $\cos^3\theta$). We shall see that the pseudoscalar ${\vec B}\cdot \curl {\vec B}$ alone determines the behavior of the pseudoscalars ${\cal{H}}^{\rm kin}$ and also of ${\cal{H}}^{\rm curr}$. The basic rotation which via $\vec g\cdot \vec\Omega$ can also form a  pseudoscalar does not play here an important role.
\begin{figure}
\vbox{
\includegraphics[width=7cm,height=5cm]{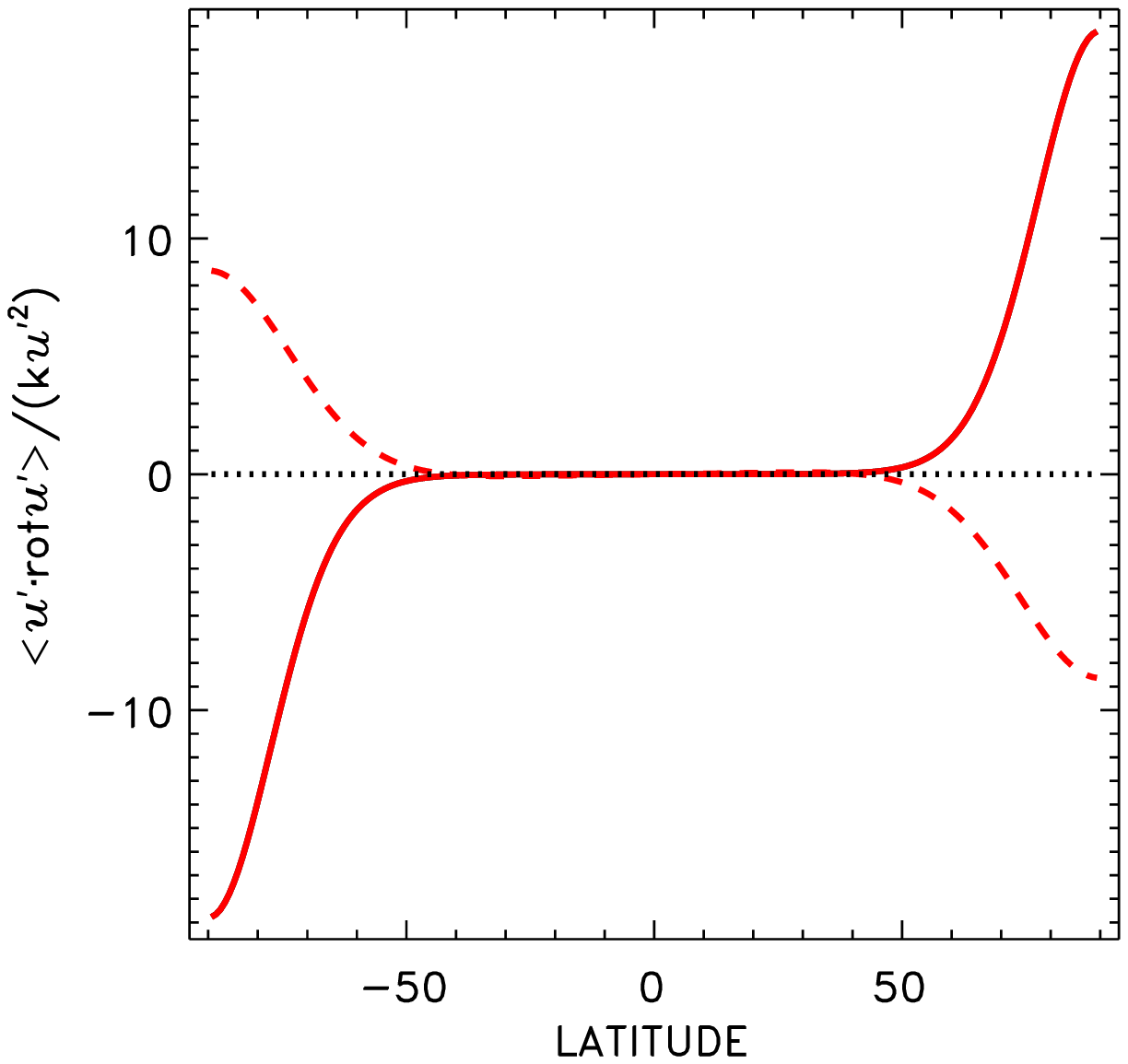}
\includegraphics[width=7cm,height=5cm]{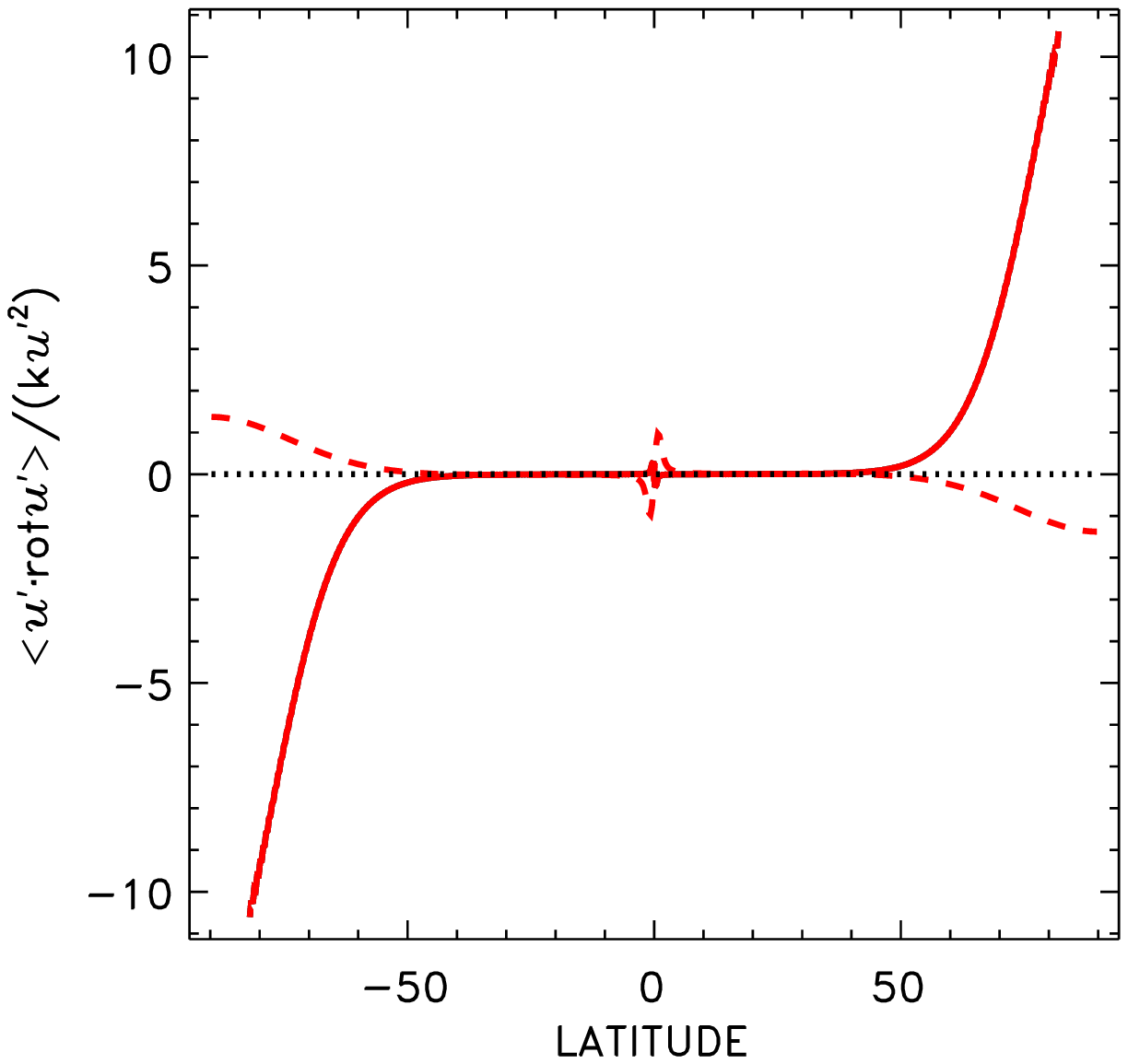}
\includegraphics[width=7cm,height=5cm]{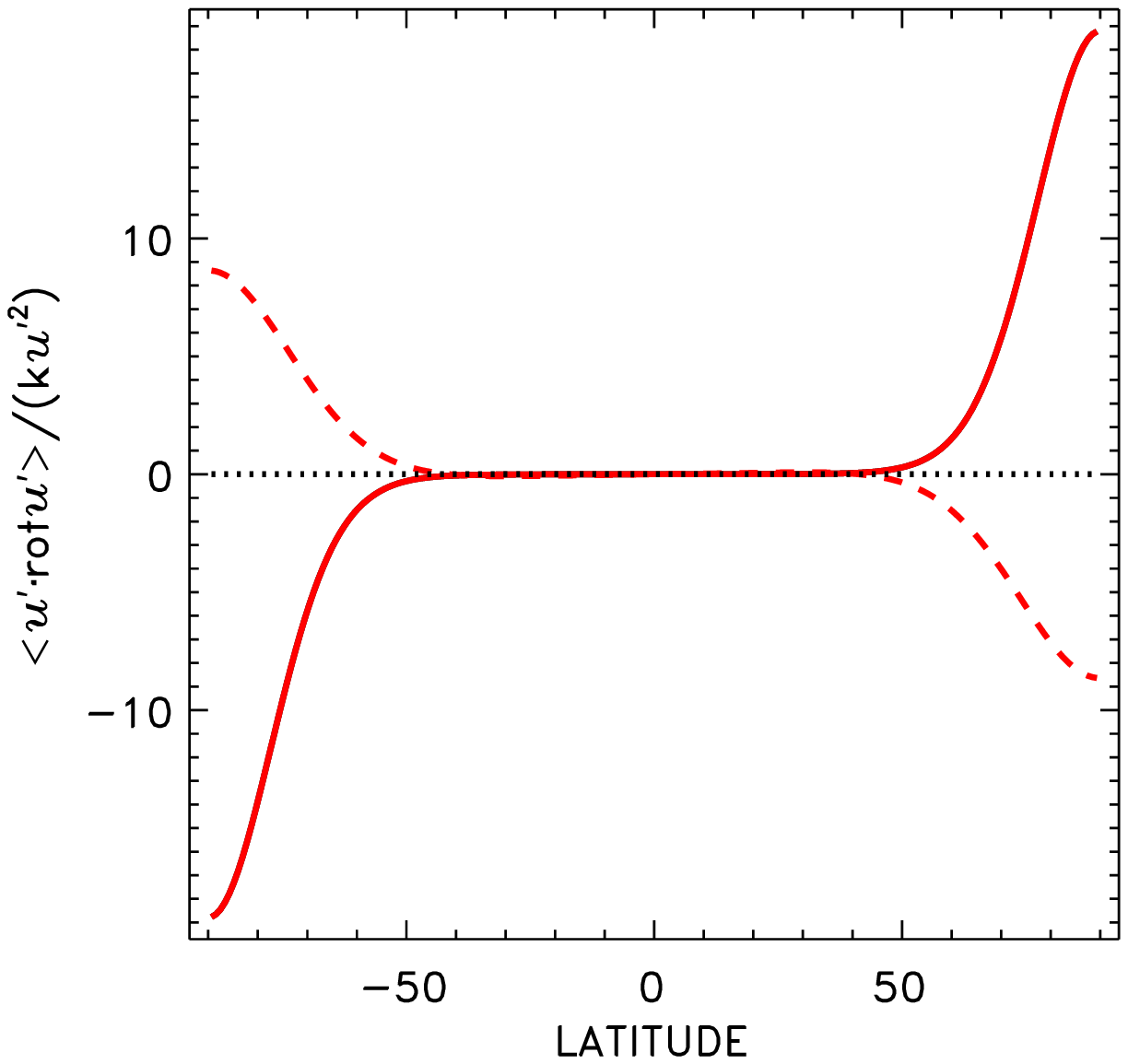}
}
\caption{Helical background field. Latitudinal profile of the  helicity   by the unstable modes S$\pm 1$ (top) and A$\pm 1$ (middle)  for  $\hat\Omega_\mathrm{A}= 0.3$  and $\hat\Omega_\mathrm{A,p} = 0.1$. Solid line: $m=1$, dashed line: $m=-1$. The net helicity does no longer vanish if both modes are excited. It is  $\beta=3N/\Omega$.
  Bottom: the S$\pm 1$ modes for $\Omega=0$.} 
  \label{fff}
\end{figure}

The ratio $\beta$ of the toroidal field amplitude and the radial field amplitude is 
\begin{equation}
\beta=\frac{B_\phi}{B_r}=\frac{\hat\Omega_{\rm A}}{\hat\Omega_{\rm A,p}} \frac{N}{\Omega}.
\label{BETA}
\end{equation}
The equation for the meridional  flow  is
\begin{eqnarray}
   &&\hat\omega\left(\hat{\cal L}V\right) =
   -\hat{\lambda}^2\left(\hat{\cal L}S\right)
   - \mathrm{i}\frac{\epsilon_\nu}{\hat{\lambda}^2}\left(\hat{\cal L}V\right)
   \nonumber \\
   &&- 2\mu\hat\Omega\left(\hat{\cal L}W\right)
   - 2\left(1-\mu^2\right)\frac{\partial\left(\mu\hat\Omega\right)}
   {\partial\mu}\frac{\partial W}{\partial\mu}
   -2 m^2\frac{\partial\hat\Omega}{\partial\mu} W
   \nonumber \\
   &&+ 2\mu\hat\Omega_\mathrm{A}\left(\hat{\cal L}B\right)
   + 2\left(1-\mu^2\right)\frac{\partial
   \left(\mu\hat\Omega_\mathrm{A}\right)}
   {\partial\mu}\frac{\partial B}{\partial\mu}
   +2 m^2\frac{\partial\hat\Omega_\mathrm{A}}{\partial\mu} B
   \nonumber \\
   &&- m\hat\Omega_\mathrm{A}\left(\hat{\cal L}A\right)
   - 2m\frac{\partial\left(\mu\hat\Omega_\mathrm{A}\right)}
   {\partial\mu} A
   - 2m\left(1-\mu^2\right)\frac{\partial\hat\Omega_\mathrm{A}}
   {\partial\mu}
   \frac{\partial A}{\partial\mu}
   \nonumber \\
   &&+ m\hat\Omega\left(\hat{\cal L}V\right)
   + 2m\frac{\partial\left(\mu\hat\Omega\right)}{\partial\mu} V
   + 2m\left(1-\mu^2\right)\frac{\partial\hat\Omega}{\partial\mu}
   \frac{\partial V}{\partial\mu}
   \nonumber \\
   &&-\frac{1}{\hat\lambda}\!\!\left(\hat\Omega_\mathrm{A,p} \left(\hat{\cal L}A\right) 
   \!+\! (1 - \mu^2) \frac{\partial\hat\Omega_\mathrm{A,p}}{\partial\mu}
   \frac{\partial A}{\partial\mu} \!-\! 
   m\frac{\partial\hat\Omega_\mathrm{A,p}}{\partial\mu} B\!\!\right)\!.
   \label{pflow}
\end{eqnarray}
The equations for the magnetic fields components are 
\begin{eqnarray}
   &&\hat\omega\left(\hat{\cal L}B\right)=
   - \mathrm{i}\frac{\epsilon_\eta}{\hat\lambda^2}\left(\hat{\cal L}B\right)
   + m\hat{\cal L}\left(\hat\Omega B\right)
   - m\hat{\cal L}\left(\hat\Omega_\mathrm{A} W\right)
   \nonumber \\
   &&- m^2\frac{\partial\hat\Omega}{\partial\mu} A
   - \frac{\partial}{\partial\mu}\left(
   \left(1-\mu^2\right)^2\frac{\partial\hat\Omega}{\partial\mu}
   \frac{\partial A}{\partial\mu}\right)
   \nonumber \\
   &&+ m^2\frac{\partial\hat\Omega_\mathrm{A}}{\partial\mu} V
   + \frac{\partial}{\partial\mu}\left(
   \left(1-\mu^2\right)^2\frac{\partial\hat\Omega_\mathrm{A}}{\partial\mu}
   \frac{\partial V}{\partial\mu}\right) 
   \nonumber \\
   &&-\frac{1}{\hat\lambda}\left(\!\hat\Omega_\mathrm{A,p} \left(\hat{\cal L}W\right) 
   \!+\! (1 \!-\! \mu^2) \frac{\partial\hat\Omega_\mathrm{A,p}}{\partial\mu}
   \frac{\partial W}{\partial\mu} - 
   m\frac{\partial\hat\Omega_\mathrm{A,p}}{\partial\mu} V\!\right)    
   \nonumber
\end{eqnarray}
and
\begin{eqnarray}
   &&\hat\omega\left(\hat{\cal L}A\right) =
   - \mathrm{i}\frac{\epsilon_\eta}{\hat\lambda^2}\left(\hat{\cal L}A\right)
   + m\hat\Omega\left(\hat{\cal L}A\right)
   - m\hat\Omega_\mathrm{A}\left(\hat{\cal L}V\right)
   \nonumber \\
   &&-\frac{1}{\hat\lambda}\!\!\left(\hat\Omega_\mathrm{A,p} \left(\hat{\cal L}V\right) 
   \!+\! (1 \!-\! \mu^2) \frac{\partial\hat\Omega_\mathrm{A,p}}{\partial\mu}
   \frac{\partial V}{\partial\mu} \!-\! 
   m\frac{\partial\hat\Omega_\mathrm{A,p}}{\partial\mu} W\!\!\right)\!.
   \label{pfield}
\end{eqnarray}

The  entropy equation
is not influenced by  the poloidal field. 

We computed the helicity   ${\cal{H}}^{\rm kin}$ of the critical modes always for the amplitude $\hat \Omega_\mathrm{A} = 0.3$ and $\hat \Omega_{\rm Ap}=0.1$  with and without basic  rotation. The  results are given in Fig.~\ref{fff}.
\begin{figure}
\vbox{
\includegraphics[width=7cm,height=5cm]{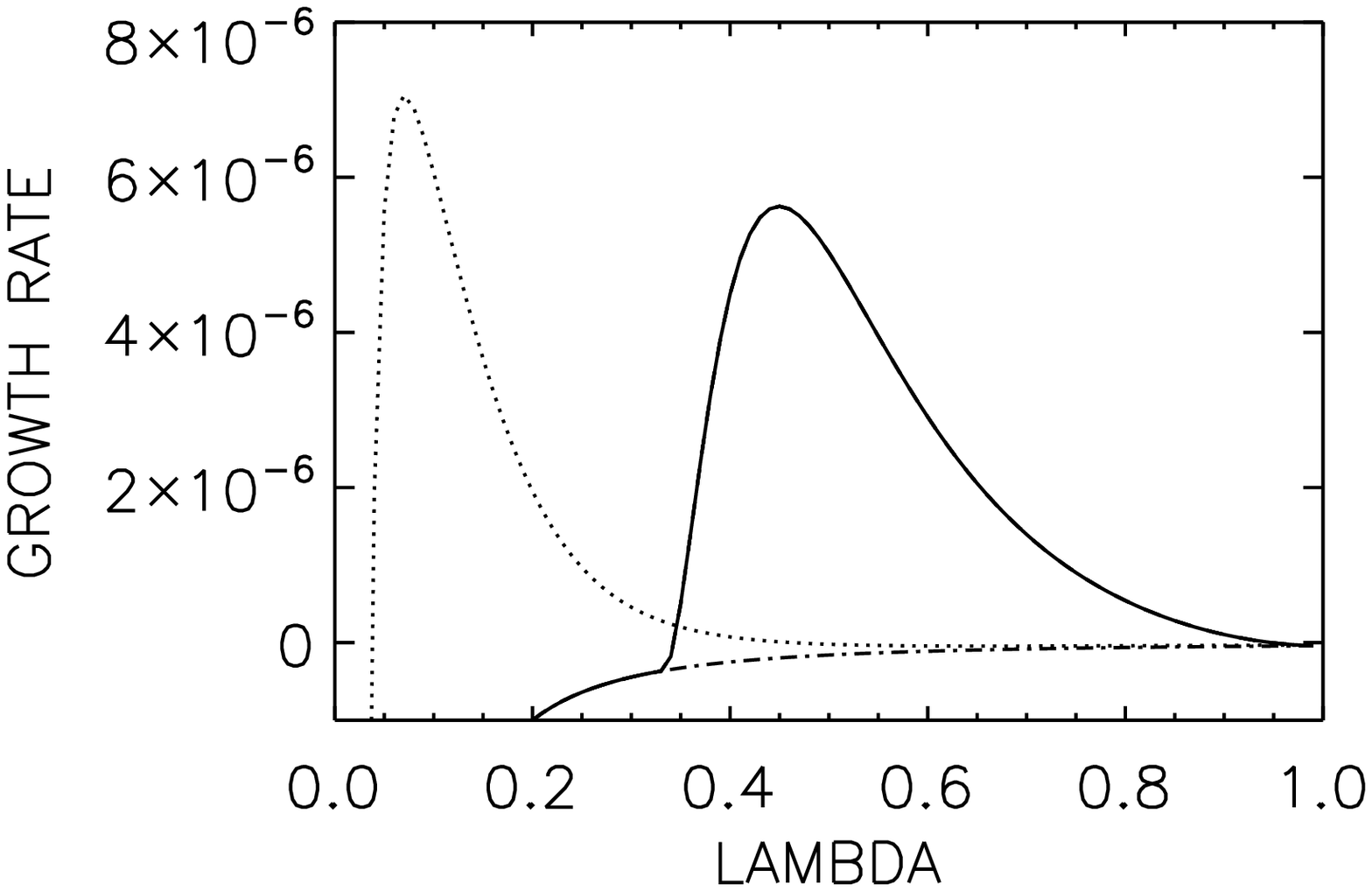}
\includegraphics[width=7cm,height=5cm]{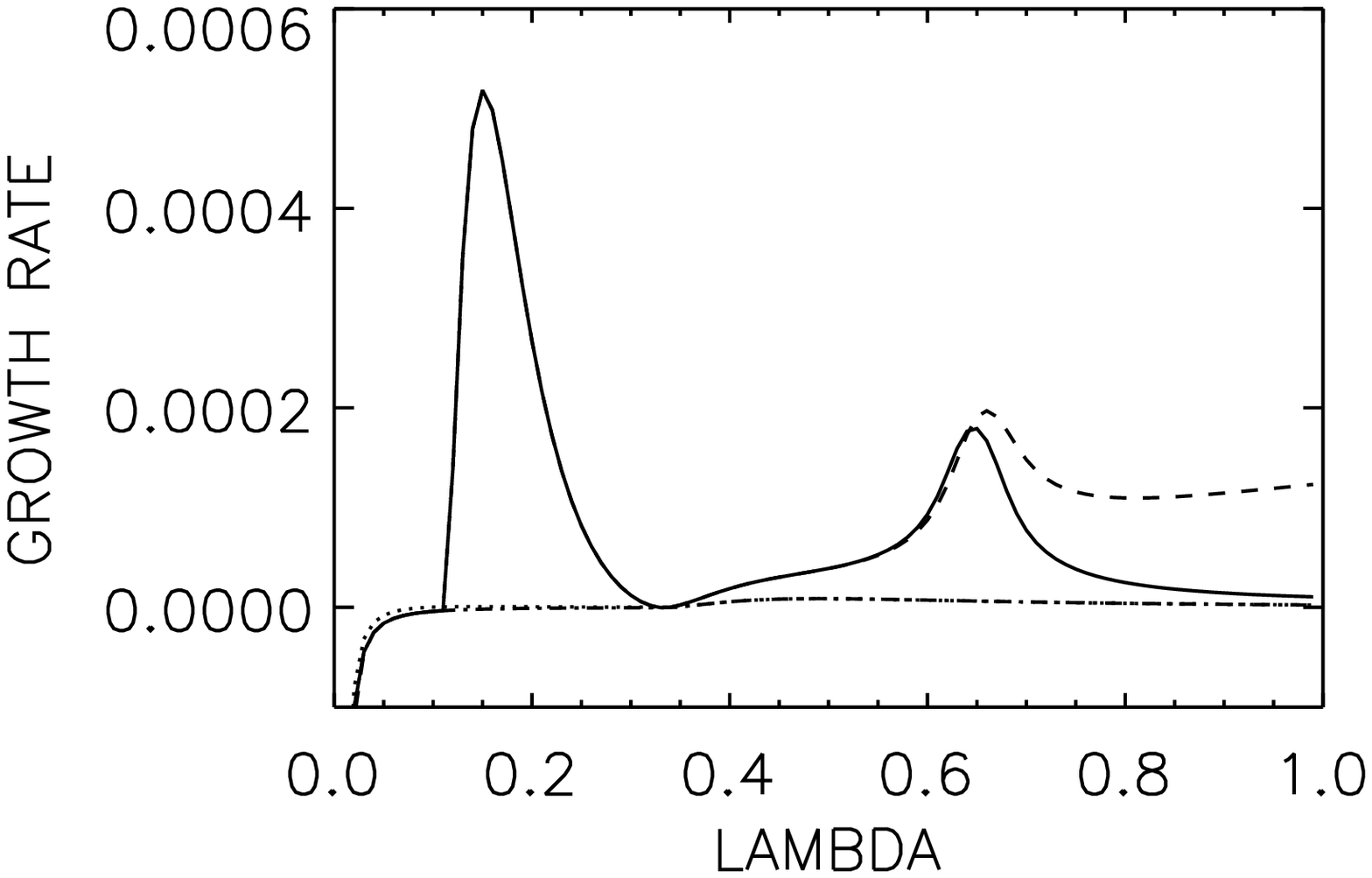}
\includegraphics[width=7cm,height=5cm]{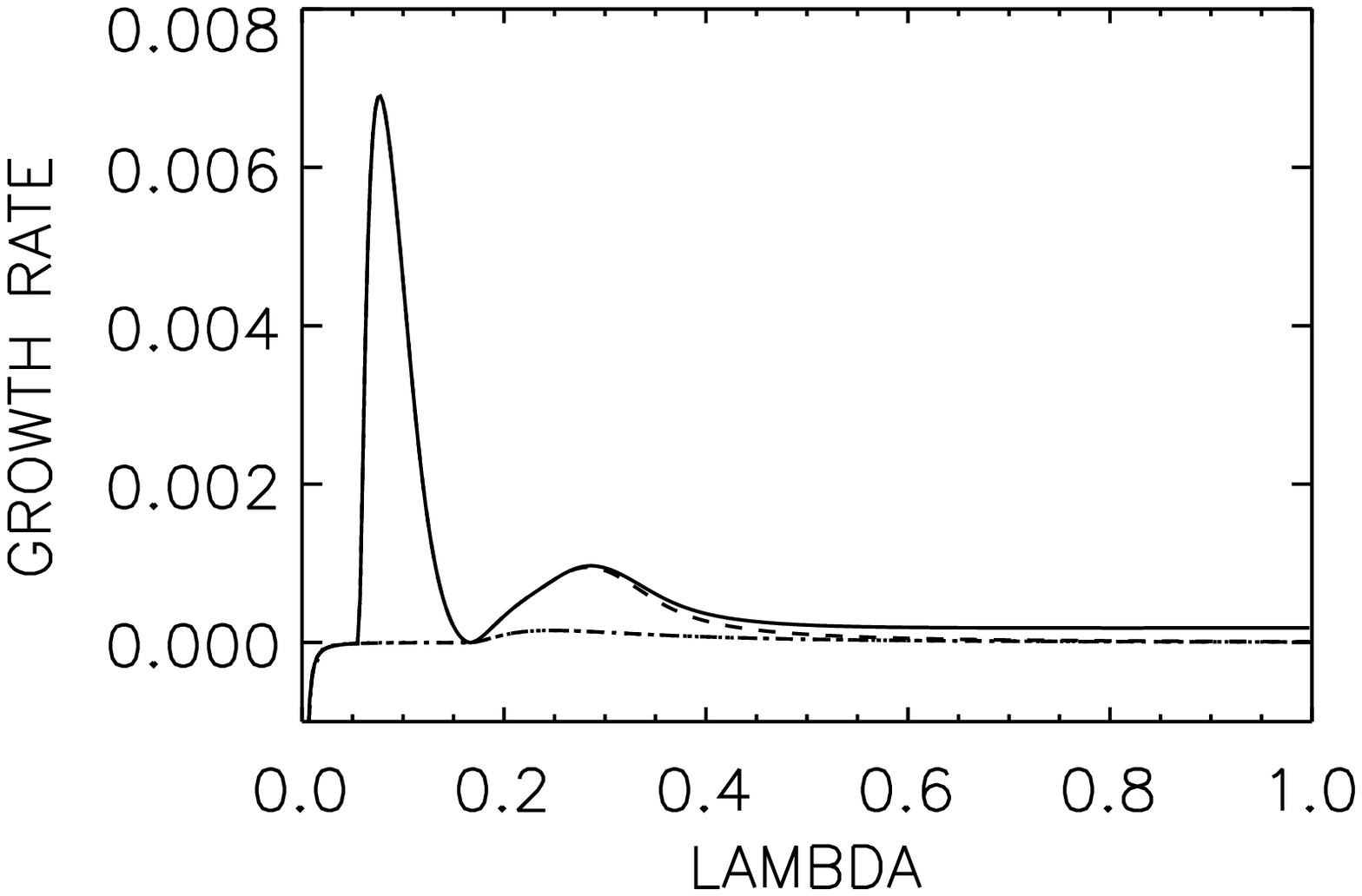}
}
\caption{The normalized  growth rates $\hat\gamma$ for the modes A$\pm 1$ and S$\pm1$   for  $\hat\Omega_\mathrm{A}= 0.3$ vs. the normalized wavelength $\hat\lambda$.  From top to bottom:  $\hat\Omega_\mathrm{A,p} = 0.3$, $\hat\Omega_\mathrm{A,p} = 0.1$, $\hat\Omega_\mathrm{A,p} = 0.05$. The modes with $m=-1$ (solid line: S-mode, dashed line: A-mode) possess  higher growth rates than the modes with $m=1$ (dash-dotted: S-mode, dotted A-mode). From top to bottom the growth rates strongly increase while the characteristic length scale decreases. The product $\hat\gamma \hat\lambda^2$  has always the value of about $10^{-6}$.} 
  \label{gr1}
\end{figure}

The solid (dashed) lines give the helicity profiles for $m=1$ ($m=-1$). Indeed, the helicity of the single modes is antisymmetric with respect to the equator (as also the current helicity of the background field). The plot at the bottom of Fig. \ref{fff} for the two modes S1 is for $\Omega=0$. Note the extremely small  differences  to  the top plot for the same modes under the  influence of   rotation. It is obviously the pseudoscalar ${\vec B}\cdot \curl {\vec B}$ directing the formation of the helicities   rather than the pseudoscalar $\vec g\cdot \vec\Omega$ formed by the global rotation.

Figure \ref{gr1} shows the normalized growth rates $\hat\gamma=\gamma/\Omega$ vs  the normalized wave length $\hat\lambda$ for three different poloidal field amplitudes. From top to bottom:  $\hat\Omega_\mathrm{A,p} = 0.3$, $\hat\Omega_\mathrm{A,p} = 0.1$, $\hat\Omega_\mathrm{A,p} = 0.05$. The peaks for the modes with $m=-1$ drift from $\hat\lambda=0.45$ to 
$\hat\lambda=0.015$. The corresponding peak values of the growth rates dramatically grow from $ 10^{-6}$ to $ 10^{-2}$ demonstrating the strong stabilization  of the Tayler instability for helical fields.  An increase of the poloidal field amplitude by a factor of 6 (from bottom to top) leads to a reduction of the growth rate by three orders of magnitude. 

The  main information of Fig. \ref{gr1} is that the growth rates of the modes with $m=1$ and $m=-1$ strongly differ. Obviously, they  always produce helicity of  opposite signs and with  different growth rates. 
The growth rate for the S1 mode with  $m=-1$ exceeds the growth rate of $m=1$
by a factor of four. It is, however, much smaller than the rotation rate. The unstable Tayler modes are thus very slow, their corresponding growth times are much longer than the  rotation periods.  The helicity by the $m=-1$ mode is {\em negative} at the north pole  opposite to the current helicity of the background field. The small-scale kinetic helicity and the large-scale current helicity   are anticorrelated (cf. Gellert, R\"udiger \& Hollerbach 2011, for a similar result in cylinder symmetry). 


\section{The alpha-effect}
Linear stability computations do not provide  the absolute value of $\alpha$. Only its latitudinal profile and its relative magnitude can be evaluated. We computed the normalized electromotive force 
\begin{equation}
  {\cal E}\ = \ 
  \frac{N}{\Omega}
  \frac{\langle {\vec u}\times{\vec b}\rangle_\phi}
  {u_{\rm rms} b_{\rm rms}} ,
  \label{10a}
\end{equation}
which in opposition to the helicity (Fig. \ref{fff}) is symmetric with respect to the equator.
$u_{\rm rms}$ is the rms velocity fluctuation after horizontal averaging (longitude and latitude), $b_{\rm rms}$ is the rms magnetic fluctuation. The factor ${N}/\Omega$ is introduced because the horizontal (velocity and magnetic) fluctuations are larger than the  radial fluctuations by just this factor. With these normalizations  the expression  (\ref{10a}) is of order  unity   independent of the actual value of $N/\Omega$. 
\begin{figure}
  \includegraphics[width=8cm,height=7cm]{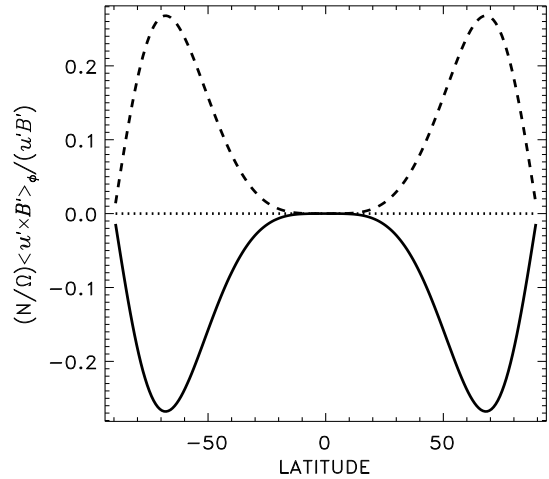}
  \caption{The same as in Fig. \ref{f10} (bottom) but for  the normalized electromotive force after Eq. (\ref{10a}) for the modes with $m=1$ (solid) and $m=-1$ (dashed). 
  $\hat\Omega_\mathrm{A,p} = 0$.
  }  
  \label{f9}
\end{figure}

Figure  \ref{f9} gives the main results  for purely toroidal fields   with  (\ref{2}). The azimuthal component of the electromotive force i) vanishes at the equator,  ii) the profile is  symmetric with respect to the equator and iii) it is highly concentrated to the poles.  
The first finding is naturally for the $\alpha$ effect but it is {\em not} for the term ${\vec \Omega}\times \vec{J}$ which could appear in the expression of the turbulence-induced electromotive force as a consequence of a rotationally induced anisotropy of the diffusivity tensor. We find, therefore,  this effect  not existing  due to the Tayler instability of toroidal fields. For more complex field pattern its existence of  cannot be excluded but it remained  small in any case.

For purely toroidal fields  even the $\alpha$ effect does not exist as the modes with opposite sign of $m$ (which have the same growth rates) do cancel each other not only with respect to their  helicities but also with respect to the resulting EMF. In the nonlinear regime a spontaneous parity breaking may happen as it has been described by Chatterjee et al. (2010) and by Gellert et al. (2011). In this case, however, it might be impossible to predict the sign and the amplitude of the $\alpha$ effect. 

The concentration of  helicity and EMF  towards the poles reflects a basic property of the Tayler instability, i.e. that the instability pattern is more present in polar regions rather than in equatorial regions (Spruit 1999; Cally 2003).

Figure  \ref{gr1}   shows the normalized growth rates $\hat\gamma$ as a function of the radial scale $\hat \lambda$  under the influence of poloidal field components.   The plots shows drastic differences of  the growth rates  between the modes of $m=1$ (dotted lines)  and $m=-1$ (solid lines). E.g., for $\hat\Omega_\mathrm{A,p} = 0.1$ both  modes with $m=-1$ possess the  maximum growth rates $\hat\gamma$ with   $5\cdot 10^{-4}$ at a wavelength of $\hat \lambda=0.15$.  The  figure shows  a very strong influence of the poloidal field amplitude on the growth rates. The growth rates are small for strong poloidal field but  they are large for weak poloidal field. We are thus confronted with the dilemma that  only background fields with finite current helicity originate fluctuations with $\alpha$ effect but  the corresponding poloidal field components     suppress the Tayler instability (see R\"udiger, Schultz \& Elstner  2011).

\begin{figure}
\vbox{
\includegraphics[width=7cm,height=5cm]{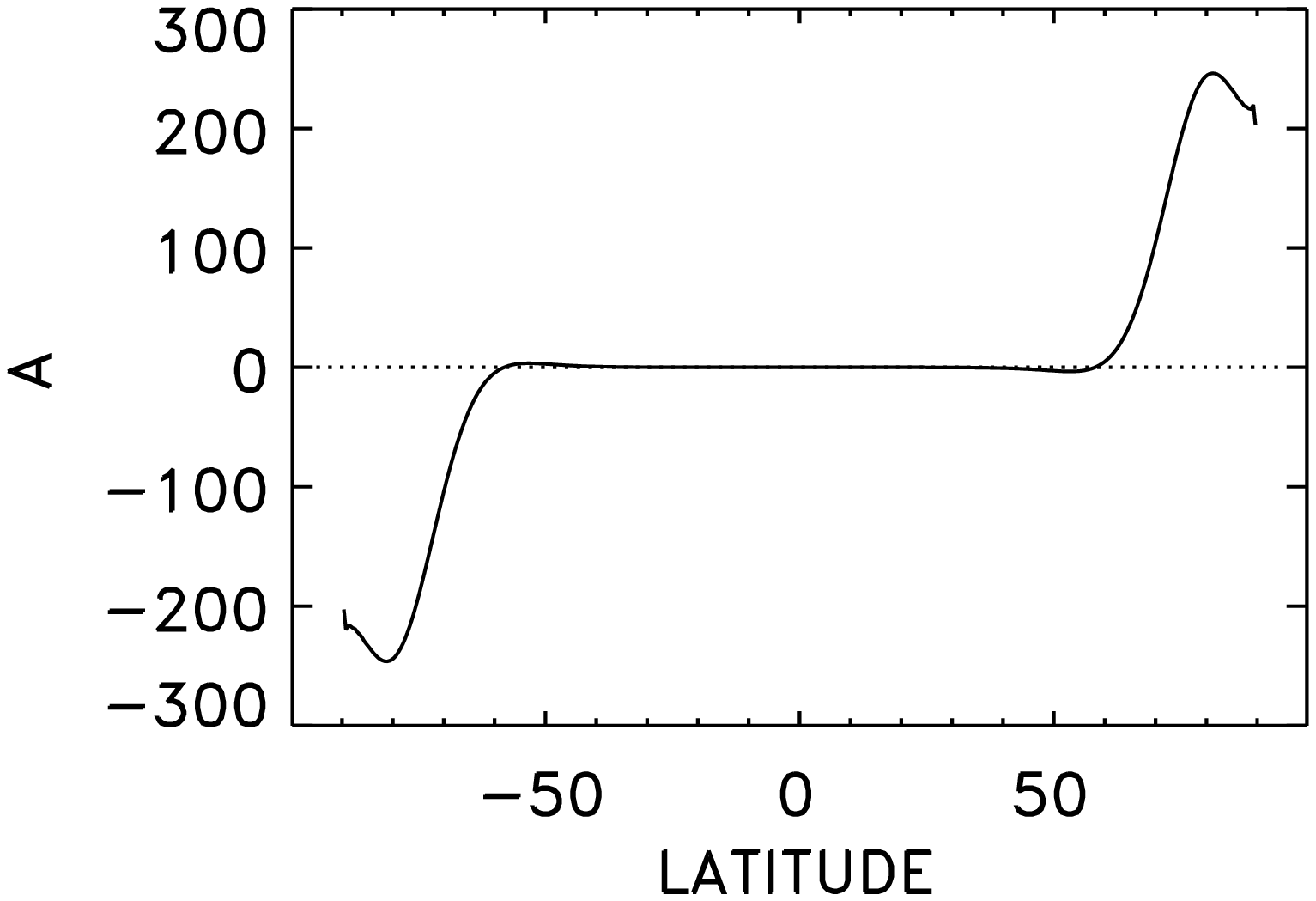}
\includegraphics[width=7cm,height=5cm]{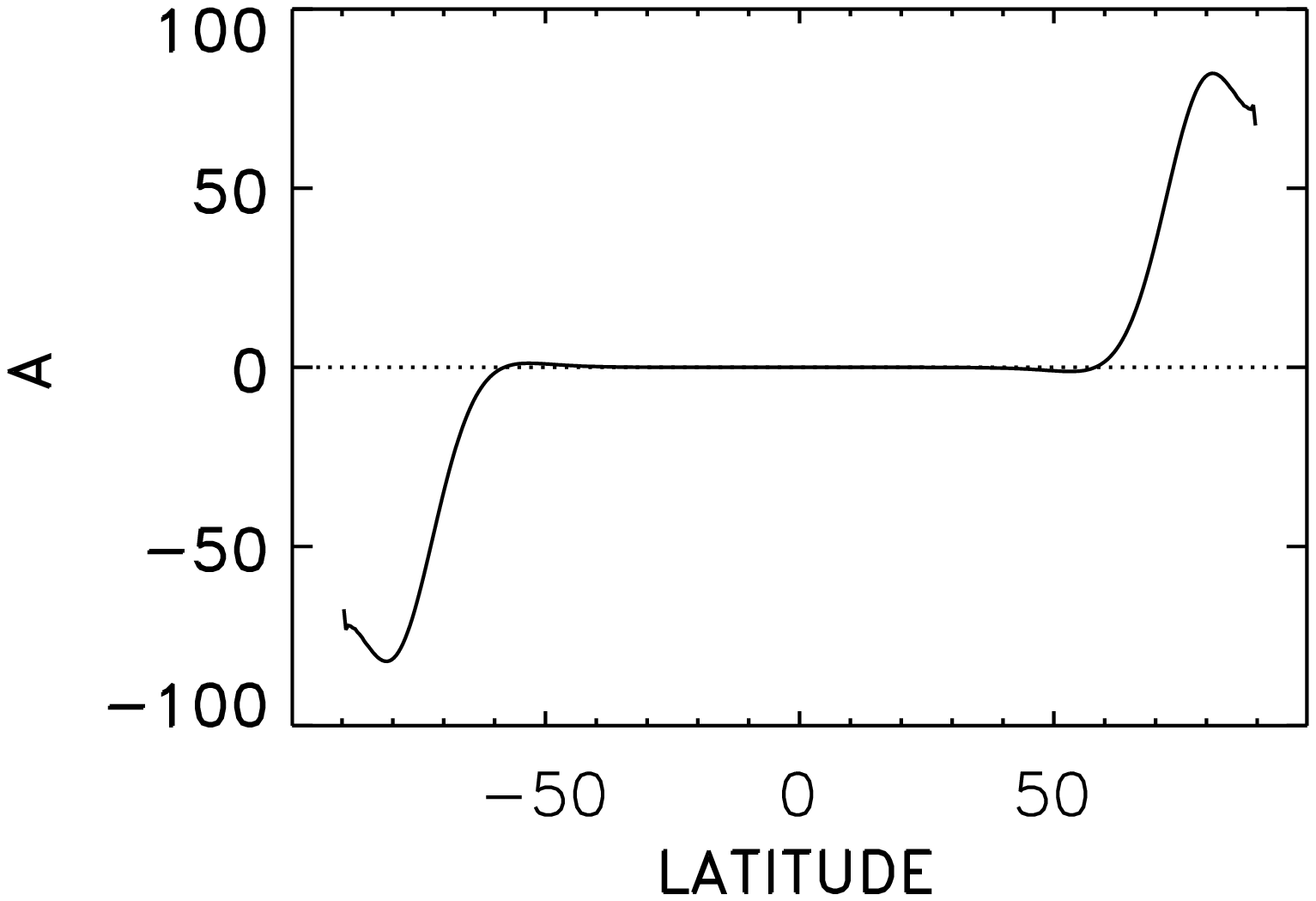}
\includegraphics[width=7cm,height=5cm]{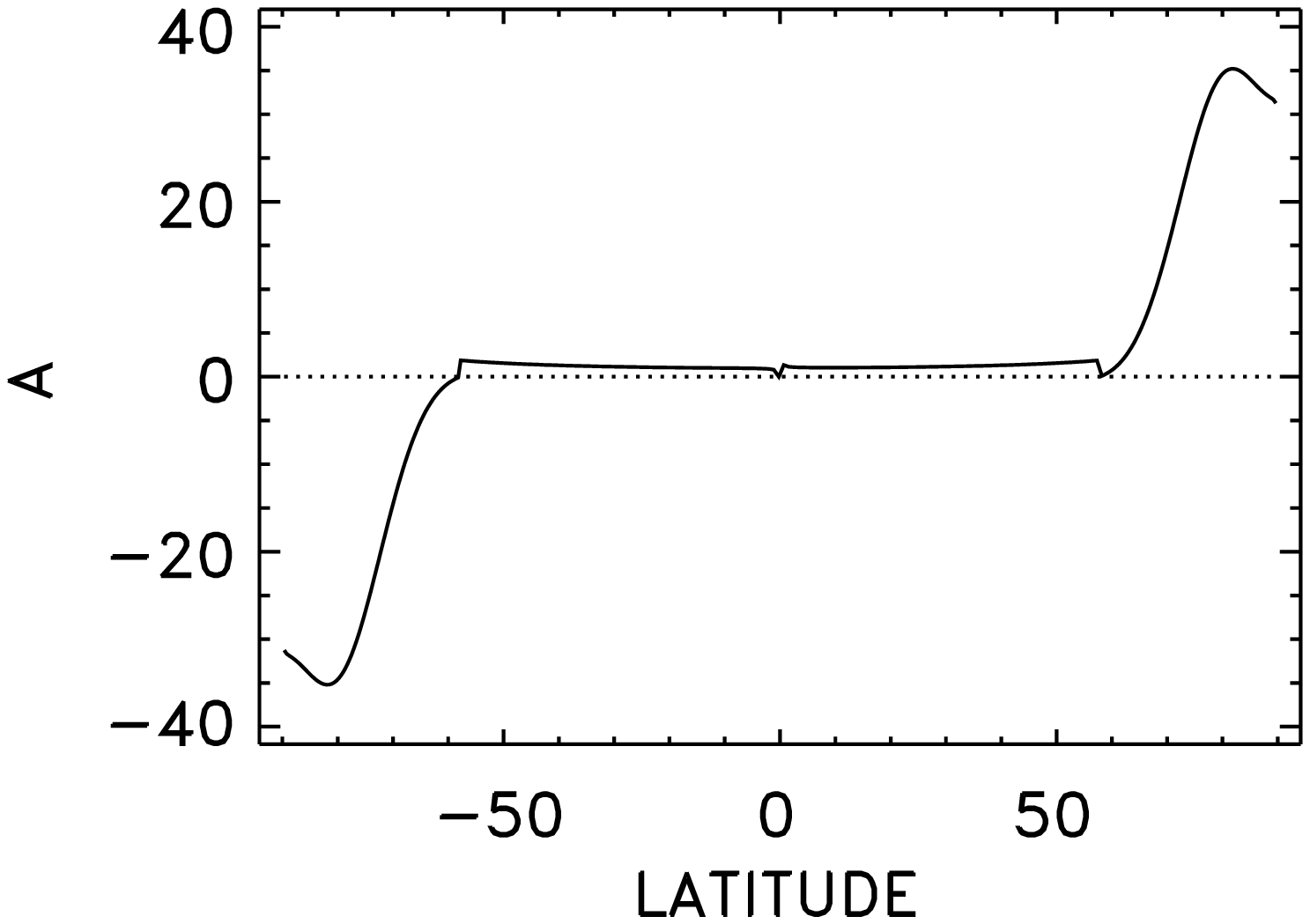}
}
  \caption{The normalized $\alpha$ effect parameter $A$ for the modes with the highest growth rates ($m=-1$, identic for A mode  and S mode).  From top to bottom:  $\hat\Omega_\mathrm{A,p} = 0.3$, $\hat\Omega_\mathrm{A,p} = 0.1$, $\hat\Omega_\mathrm{A,p} = 0.05$. Note i) the extreme concentration of $A$ to the poles and ii) the $A$ amplitudes strictly running with $\hat\Omega_\mathrm{A,p}$. For the given magnetic geometry the $\alpha$ effect is always positive (negative) at the northern (southern) hemisphere.  $\hat\Omega_\mathrm{A}= 0.3$.
} 
  \label{f2}
\end{figure}

For the further discussion the quantity
\begin{equation}
  A\ = \ 
  \frac{{\cal E}}
  {\sin\theta\cos\theta} = \frac{\alpha B_0}{u_{\rm rms}b_{\rm rms}}\frac{N}{\Omega} 
  \label{10}
\end{equation}
is introduced which is antisymmetric with respect to the equator as expected for the (normalized) $\alpha$ effect. The $\sin\theta\cos\theta$ in the denominator eliminates the latitudinal profile of   the toroidal field   and  $B_0 = \sqrt{\mu_0\rho}\ r\Omega_\mathrm{A}$ is its amplitude (see Eq.  (\ref{2})).

Also the  resulting $A$-profiles   are highly concentrated to the  poles (Fig.~\ref{f2}). 
The  plots show the profiles for the  weak-field cases $\hat\Omega_\mathrm{A} = 0.3$ with  $\hat\Omega_\mathrm{A,p} = 0.3$, $\hat\Omega_\mathrm{A,p}=0.1$ and $\hat\Omega_{\rm A,p}=0.05$.  The modes with  the fastest growth produce an   $\alpha$ effect which is always positive (negative)  for positive (negative) current helicity at the northern (southern) hemisphere of the background field. The result is an  $\alpha$ effect anticorrelated with the small-scale helicity and positively correlated with the large-scale  pseudoscalar ${\vec B}\cdot \curl {\vec B}$. Exactly the same  relations have been derived by nonlinear simulations of the kink-type instability  for an incompressible fluid in a cylindric setup by Gellert et al. (2011). We are thus encouraged to favor the results for the modes with the highest growth rates.

For very weak poloidal field ($\hat\Omega_{\rm A,p}=0.01$, not shown)  the  $\alpha$ effect is already so  small (and its sign fluctuates) that the $\Omega_{\rm A,p}=0.01$ seems to form the lower limit of the helicity production by poloidal fields.

In the weak-field regime ($\hat\Omega_\mathrm{A} < 1$) the instability excites stronger  magnetic fluctuations rather than flow fluctuations. The ratio of fluctuating Alfv\'en velocity $v_{\rm rms} = b_{\rm rms}/\sqrt{\mu_0\rho}$ to $u_{\rm rms}$ results as  $v_{\rm rms}/u_{\rm rms} \simeq 470$ for $\hat \Omega_\mathrm{A} = 0.01$ and $v_{\rm rms}/u_{\rm rms} \simeq 46$ for $\hat \Omega_\mathrm{A} = 0.1$. Hence, we find  
\begin{equation}
\frac{v_{\rm rms}}{u_{\rm rms}} \simeq \frac{4.7}{\hat\Omega_\mathrm{A}}
 \label{ratio}
\end{equation}
 for weak fields. For strong fields ($\hat\Omega_\mathrm{A} > 1$) the fluctuations become close to equipartition, i.e. $v_{\rm rms} \simeq u_{\rm rms}$.

\section{Dynamo theory}
On the basis of the obtained informations about the $\alpha$ effect (amplitude and latitudinal profile)  we have to probe  the possible existence of a dynamo mechanism. 
\subsection{Eddy diffusivity}
We start to estimate  the typical velocity perturbation and the eddy magnetic-diffusivity. We write $u\simeq \lambda/\tau$ and use the growth time $\tau_{\rm gr}=1/\gamma$ as the timescale $\tau$. Hence,
\begin{equation}
u\simeq \gamma \lambda= \pi \hat\gamma \hat\lambda \frac{\Omega^2 r}{N}.
\label{uuu}
\end{equation}
The first step  holds for purely toroidal fields ($\Omega_{\rm A,p}=0$). From Fig.~\ref{f10} (top) one finds $\hat\gamma \hat\lambda\equiv 10^{-4}$ so that with solar values $u\simeq 1$ mm/s results. The estimate $\eta_{\rm T}\simeq  u \lambda$  leads to $\eta_{\rm T}\simeq 10^5$ cm$^2$/s as the order of magnitude of the eddy diffusivity. The value increases for fast rotating giants by one or two orders of magnitudes. It perfectly fits the diffusion coefficients for chemicals which are needed to explain the weak lithium depletion of solar type stars (Barnes, Charbonneau \& MacGregor 1999).

The same estimate $\eta_{\rm T}\simeq \gamma \lambda^2$ yields
\begin{equation}
\eta_{\rm T}\simeq \pi^2 \hat\gamma \hat\lambda^2 \frac{\Omega^2 }{N^2} \Omega r^2.
\label{eta}
\end{equation}
The three examples for helical background fields given in Fig. \ref{gr1} lead to a common value of  $\hat\gamma \hat\lambda^2\simeq 10^{-6}$. Hence,   $\eta_{\rm T}\simeq 3 \cdot 10^5$ cm$^2$/s results for the upper radiation zone of the Sun. This value is very close to the above estimation for purely toroidal fields. It is increased by more than  three orders of magnitudes if more massive stars are considered.  Figure \ref{f5} shows the details of a model with a mass of  $3M_\odot$ and 10 days rotation period.

\subsection{Dynamo conditions}
The $\alpha$ effect enters the dynamo theory via the dimensionless dynamo number
\begin{equation}
C_\alpha= \frac{\alpha\ r}{\eta_{\rm T}}.
\label{Calf}
\end{equation}
 Equation  (\ref{10}) together with the heuristic relation $u^2/\eta_{\rm T}\simeq \gamma$ directly  provides  
\begin{equation}
  C_\alpha\ = \ \frac{5}{\hat\Omega^2_{\rm A}}
  \frac{\Omega}{N}\  {\hat\gamma} \ A.
  \label{Calpha}
\end{equation}
Hence, the product $\hat\gamma A$ determines the effectivity of the $\alpha$ effect. With $\hat\gamma\ A\simeq 0.03$ a value of $C_\alpha\simeq 10^{-(2...3)}$ results for the solar model. Note that the buoyancy frequency linearly enters the equation. Also for the $3M_\odot$ star of Fig. \ref{f5} the $C_\alpha\simeq 0.1$ remains small. The operation of a (stationary)  $\alpha^2$ dynamo is thus excluded for radiation stellar zones. The  existence of an (oscillating) $\alpha\Omega$ dynamo, however,  remains possible. All $\alpha\Omega$ dynamos can work with very small $\alpha$ effect if only $C_\Omega$ 
\begin{equation}
C_\Omega= \frac{ \Omega_0\ r^2}{\eta_{\rm T}}.
\label{COm}
\end{equation}
is high enough and this is always possible for sufficiently small  eddy diffusivity $\eta_{\rm T}$. The   only consequences of very  small $\alpha$ effect are i) that the ratio of toroidal and poloidal magnetic field components becomes very large and ii) also the growth time of the dynamo instability becomes  large. The growth time  of the weakly supercritical $\alpha\Omega$ dynamo for  $\eta_{\rm T}\simeq 10^5$cm$^2$/s is extremely long (order of Gyrs for the Sun). With (say) 3\% differential rotation this value of the eddy diffusivity leads for the Sun to $C_\Omega$ of order 10$^7$.

There is another   possibility to proceed. For all $\alpha\Omega$ dynamos the ratio $\beta$ of the toroidal and the radial field amplitudes is 
\begin{equation}
\beta\simeq \epsilon \sqrt{\frac{C_\Omega}{C_\alpha}}.
\label{beta}
\end{equation}
 The scaling parameter $\epsilon$ is about 0.05 (see Table \ref{tab}).
Hence, $C_\Omega\simeq \beta^2 C_\alpha/\epsilon^2$. The excitation condition for such dynamos can be written as 
\begin{equation}
C_\alpha C_\Omega= D_{\rm crit},
\label{Ccrit}
\end{equation}
 where $D_{\rm crit}$ runs inversely with the shear.  With (\ref{beta}) $ C_\alpha \gsim \sqrt{\epsilon^2 D_{\rm crit}}/\beta$ results as excitation condition. 
One finds
\begin{equation}
\hat\gamma A \gsim \frac{\epsilon}{5}\sqrt{ D_{\rm crit}} \hat\Omega_{\rm A,p} \hat\Omega_{\rm A}
\label{AA}
\end{equation}
for fixed shear. Note that the condition (\ref{AA}) does no longer contain the stellar parameters like $\Omega/N$ and/or the radius. The following findings are thus valid for all stellar radiation zones.

Our first example is formed by $\hat \Omega_{\rm A}=0.3$ and $\hat \Omega_{\rm A,p}=0.3$.  
The condition (\ref{AA})  then  reads $\hat\gamma A\gsim 0.06\epsilon  \sqrt{D_{\rm crit}}$. As we shall see below, a typical value for the dynamo number is $\epsilon\sqrt{ D_{\rm crit}}\simeq 30$ so that  for dynamo excitation  \begin{equation}
\hat\gamma A\gsim 2 \hat\Omega_{\rm A,p}.
\label{cond}
\end{equation}
This relation must be read as an  equation for the supercritical value of $\hat \Omega_{\rm A,p}$.  The excitation is thus formally more easy for smaller  poloidal field. After the numerical results in the Figs. \ref{gr1} (top) and \ref{f2} (top)  one finds  
 $\hat\gamma A\simeq 10^{-3}$ for $\hat \Omega_{\rm A,p}=0.3$  which is by far too small. 
For  $\hat \Omega_{\rm A,p}=0.1$   it is  $\hat \gamma A\simeq 0.03$  which is also not supercritical.  The comparison of the Figs. \ref{gr1} and \ref{f2} demonstrates that for large poloidal fields the $A$ grows but in the same time the growth rate drastically sinks which indicates the stabilizing action of the poloidal fields. An $\alpha\Omega$ dynamo in radiation zones cannot work, therefore,  with too strong poloidal fields. 

Note that with $\hat\gamma A\simeq 0.2$ for $\hat \Omega_{\rm A,p}=0.05$  the condition (\ref{cond}) can  be fulfilled due to the increase of the growth rate. For smaller  $\hat \Omega_{\rm A,p}$ the ability of the poloidal field to create a coherent  $\alpha$ effect sinks. For $\hat \Omega_{\rm A,p}=0.01$  the function $A$ has already two signs at either hemisphere. 
However, as the condition (\ref{cond}) can be fulfilled  the    existence of an $\alpha\Omega$ dynamo cannot  be excluded by the  numerical results. We must thus solve the dynamo equations in order to probe the existence of such dynamos which work with very small values of the $\alpha$ effect. 

 
The $\alpha$ effect plotted in Fig. \ref{f2}  shows another complicating characteristics, i.e. its concentration at the poles. 
In the next Section nonlinear $\alpha\Omega$ dynamo models are thus constructed with a weak  solar-type latitudinal differential rotation and an $\alpha$ effect which is concentrated at the poles. It is known from the theory of $\alpha\Omega$ dynamos operating with $\cos\theta$ profiles of the $\alpha$ effect that they produce  too polar butterfly diagrams. We must  thus expect that  the existence of $\alpha\Omega$ dynamos to produce mid-latitude  belts like in Eq.~(\ref{2}) with pole-concentrated $\alpha$ effect is unlikely.

\subsection{Dynamo models}
The influence of the polar concentration of magnetic-induced  $\alpha$ effect on a possible $\alpha\Omega$ dynamo can easily be  probed.     The stability   of an axisymmetric toroidal background field has been considered so that only  the numerical values of an axisymmetric $\alpha$ are known. This scenario is only  consistent  if the  possible $\alpha\Omega$ dynamo produces  axisymmetric toroidal field belts at the  latitude of the original background field. We  know that this is true for an $\alpha$ effect (plus solar-type differential rotation) with the standard $\cos \theta$-profile. We have to check, therefore,  whether this result remains true for the $\alpha$ effect concentrated close to the poles.

To this end a simple  model is constructed. In a   spherical shell between the normalized radii  0.6  and  1 
the  rotation frequency  is 
$\Omega=C_\Omega(1 - 0.03 \cos^2\theta)$. The small value of the relative shear allows to consider the fluid as hydrodynamically stable (see Watson 1981). Possible production of hydrodynamic-induced $\alpha$ effect in radiative zones such as that by Dikpati \& Gilman (2001) is therefore excluded. 
 The shear peaks at 45$^\circ$ which is the same latitude as that of the peak of the toroidal field. 
To model the polar concentration of the magnetic-induced $\alpha$ effect  the expression  
\begin{equation}
 \alpha={C_\alpha}\ \cos^\ell \theta
 \label{ell}
 \end{equation}
  is used with the  free parameter $\ell$ fixing the latitudinal profile of the $\alpha$ effect. The polar concentration  presented in Fig. \ref{f2} leads to rather high values of $\ell$.
A perfect-conductor boundary condition is used  at the inner radius.  The 
outer  computing domain is extended to the radius 1.2. Outside the stellar surface 
the diffusivity is increased by a factor of 10. For the outermost  boundary a 
 pseudovacuum condition is used. The method of a global quenching of the $\alpha$ effect is used
in order to find the characteristic eigenvalue of marginal dynamo instability (Elstner, Meinel \& R\"udiger 1989).

 Table \ref{tab} presents the numerical results for a reference dynamo model with  3\% differential rotation and for $C_\Omega=257,000$. Only the  numbers are given for marginal dynamo instability. If the real  $C_\alpha$ differs from the reference $C_\alpha$ by a factor $1/\gamma$ then the resulting $\beta$ also differs by the factor $\gamma$, so that
 \begin{equation}
 \beta C_\alpha = \beta^{\rm ref}  C^{\rm ref}_\alpha\simeq 30.
 \label{ref}
 \end{equation}
This condition is automatically fulfilled for all dynamo models which fulfill the excitation condition (\ref{AA}) independent on the particular stellar model. 
All the dynamos listed in Table \ref{tab} provide positive (negative) values of 
the large-scale current helicity ${\vec B}\cdot \curl {\vec B}$ at the northern (southern) hemisphere. As we have assumed the same constellation for the above  background field producing the $\alpha$ effect the theory is thus consistent under this aspect.

We have also  to probe whether the star rotates fast enough to produce sufficiently large  $C_\Omega$ for the required dynamo numbers of the reference values of Table \ref{tab}. It is 
\begin{equation}
 C_\Omega = \frac{D_{\rm crit}}{C_\alpha}= \frac{\sqrt{D_{\rm crit}}}{\epsilon} \frac{\hat\Omega_{\rm A}}{\hat\Omega_{\rm A,p}} \frac{N}{\Omega}.
 \label{Com}
 \end{equation}
Note at first that the excitation is much more easy for massive stars as their $N/\Omega$ is smaller (Fig. \ref{f5}). Inserting the characteristic numbers of our models the result is $C_\Omega \gsim  10^5\ N/\Omega$ which is fulfilled by both the considered stellar models by their estimated values $C_\Omega\simeq 10^{10}$. Either the Sun as the considered hot stars with rotation periods of a couple of days rotate fast enough the excite an $\alpha\Omega$ dynamo with a very weak shear (3\%).

To discuss the influence of the  quantity $\ell$ we present models with $\ell=1,\ \  \ell=5$  and $\ell=15$. The latter value describes the  strongest concentration of the $\alpha$ effect to the poles. For $\ell=1$ (i.e. $\alpha \propto \cos\theta$) the dynamo-generated toroidal field peaks exactly at $\theta=45^\circ$ so that the presented stability analysis which bases on a toroidal field which also peaks at $\theta=45^\circ$ (see Eq. (\ref{2})) would be  consistent. This case together with the excitation condition (\ref{cond}) which is fulfilled by the magnetic amplitudes $\hat\Omega_{\rm A}=0.3$
and  $\hat\Omega_{\rm A,p}=0.05$ would strongly suggest the existence of an $\alpha\Omega$ dynamo in magnetized radiation zones of hot stars.

\begin{figure}
  \resizebox{\hsize}{!}{\includegraphics{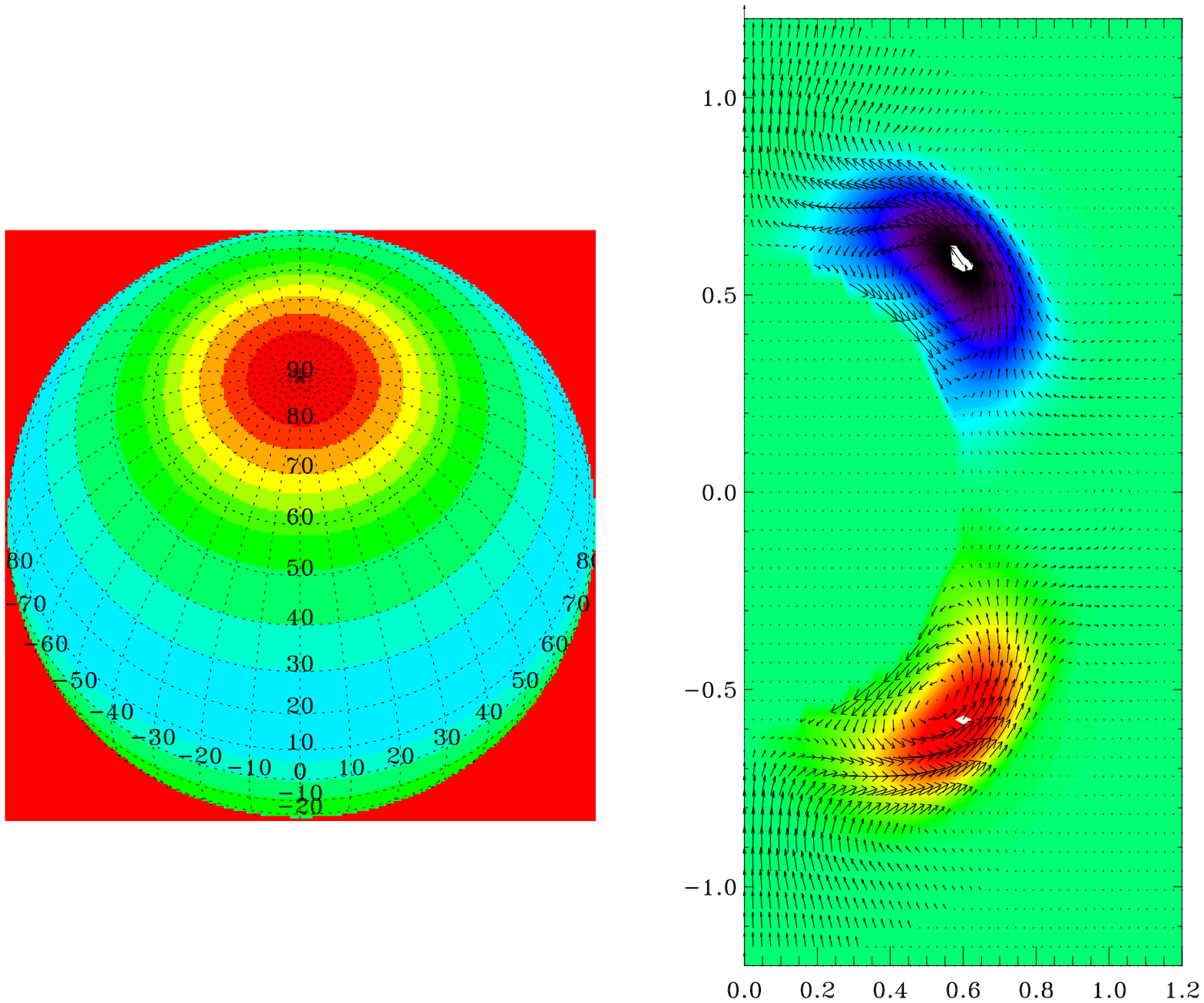}}
  \vspace{0.1 truecm}
  \resizebox{\hsize}{!}{\includegraphics{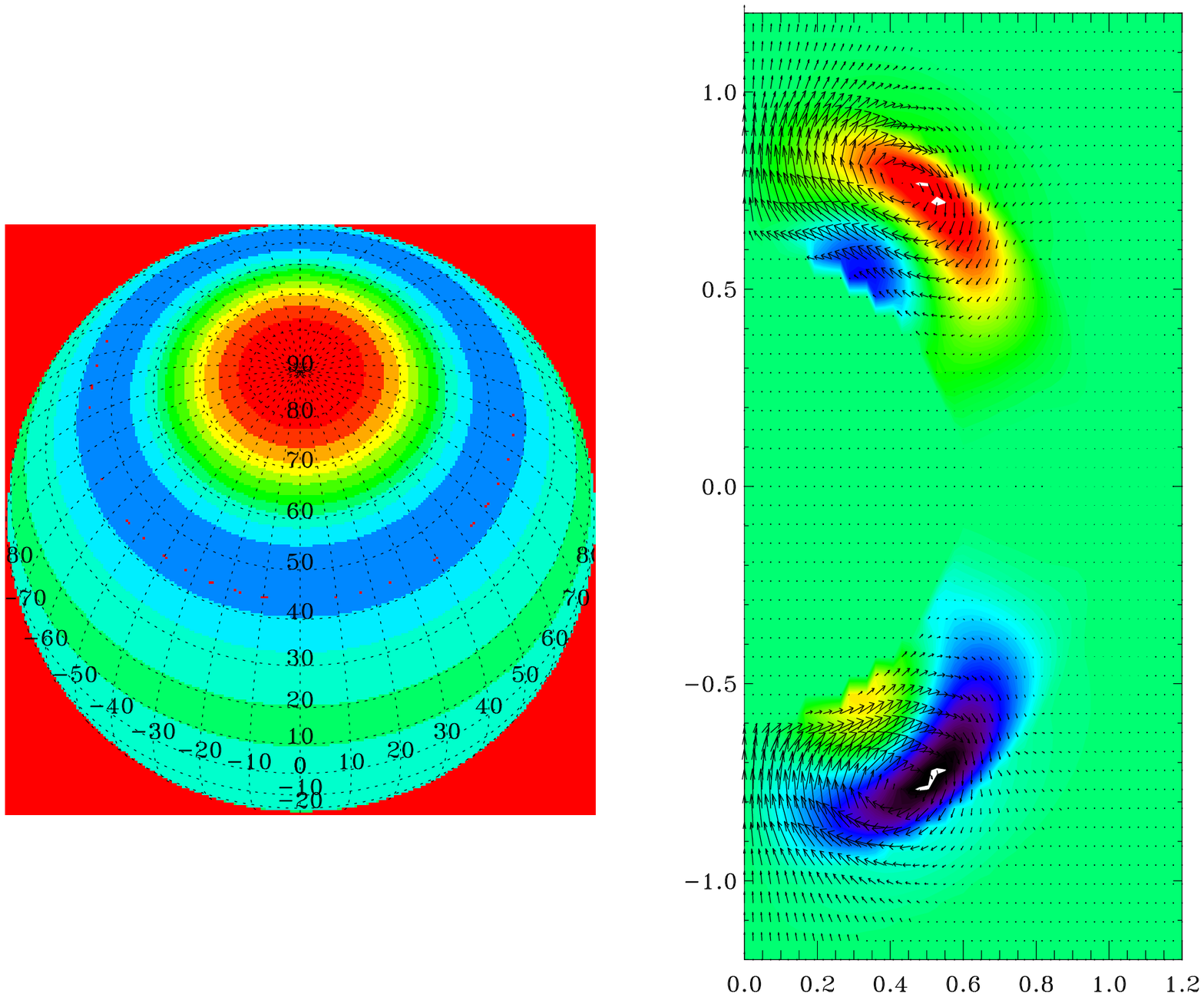}}
  \vspace{0.1 truecm}
  \resizebox{\hsize}{!}{\includegraphics{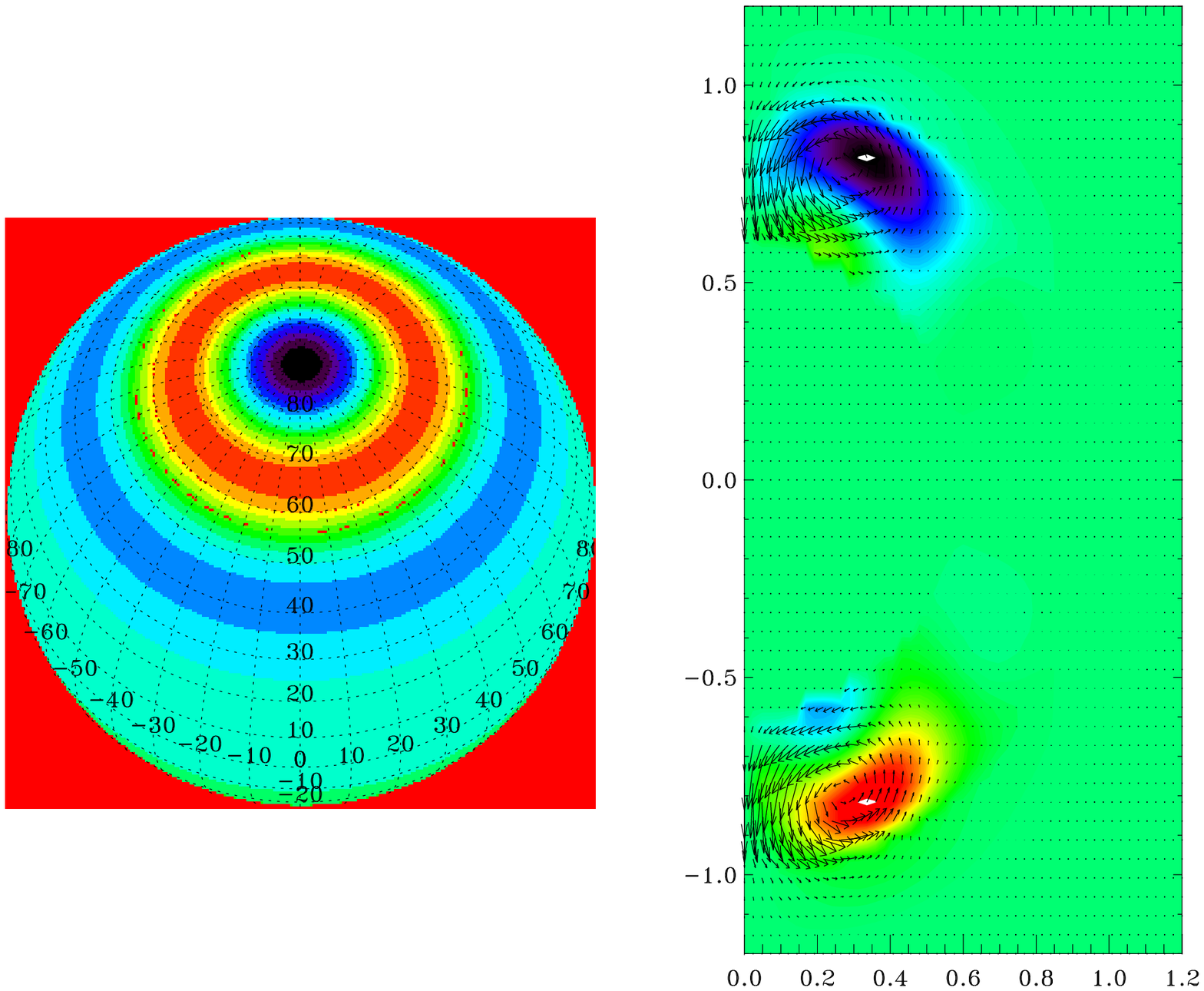}}
  \caption{The  field geometry of $\alpha\Omega$ dynamo models with $\ell=1$ (top), $\ell=5$ (middle), $\ell=15$ (bottom). Left:  radial field at the surface. Right: toroidal field pattern. The magnetic field belts are drifting radially  during the cycle.}  
  \label{f6}
\end{figure}

However, the models listed in Table \ref{tab} provide the  result that  the induced toroidal field belts become more and more concentrated to the poles for increasing $\ell$. This behavior is insofar not trivial as the generation of the toroidal field by the differential rotation does {\em not} depend on  $\ell$. 
Such  polar belts resulting for $\ell>1$ can never reproduce the  mid-latitudinal belts of Eq. (\ref{2}) on which  the $\alpha$ effect bases. The first example of Fig.~\ref{f6} (top) is the  standard $\alpha\Omega$ dynamo with the latitudinal profile $\alpha \propto \cos\theta$. It produces axisymmetric toroidal magnetic belts of dipolar symmetry at the same  latitude where the shear $\partial \Omega/\partial \theta$ has a maximum. For growing $\ell$  the belt position drifts more and more polewards (the field position becomes $45^\circ > \theta > 135^\circ$). It is thus {\em not possible}  to maintain toroidal field belts in mid-latitudes if the $\alpha$ effect mainly exists  in the polar region ($\ell > 1$).

\begin{table}
\caption{\label{tab} Reference  dynamo models   with 3\% differential rotation and for $C_\Omega=257,000$. The last column gives the polar distance of the maximum position of the northern toroidal field belt. }
\begin{tabular}{ccccccc}
\hline
\hline
$\ell$& $D_{\rm crit}$ &$C_\alpha$ &$\beta$ & $\epsilon$ & $\epsilon\sqrt{D_{\rm crit}}$  & $\theta$\\
\cline{1-7}\\
1 & 118,000 &0.46 &56 & 0.07 & 26 &45$^{\circ}$\\
5 & 347,000 & 1.35&23 & 0.05 & 31 &38$^\circ$\\
15 & 976,000 &3.8 & 6.3 & 0.02 & 24&23$^\circ$\\
\hline
\end{tabular}
\end{table}

\section{Conclusions}

It is demonstrated with a linear theory that the Tayler instability of a toroidal magnetic field in a density-stratified radiation zone of hot stars does not produce helicity and/or $\alpha$ effect. If, however, a weak poloidal field component is added forming a large-scale current helicity ${\vec B} \cdot {\rm curl} {\vec B}$ then the instability leads to small-scale helicity, current helicity and also to $\alpha$ effect. If ${\vec B} \cdot {\rm curl} {\vec B}$ is {\em positive} in the northern hemisphere then the helicity $\langle {\vec u}\cdot {\rm curl} {\vec u} \rangle$ is {\em negative} at the northern hemisphere if only  the modes are considered which grow fastest. Hence, the small-scale kinetic helicity and the large-scale current helicity of the background field are anticorrelated.

As it is often the case, the corresponding $\alpha$ effect is anticorrelated with the kinetic helicity and, therefore, positively correlated with the large-scale current helicity ${\vec B} \cdot {\rm curl} {\vec B}$. The calculations lead to two basic properties of this $\alpha$ effect. It is i) small, i.e. the normalized value $C_\alpha$ is only of order $10^{-(2\dots 3)}$ and ii)  concentrated at the poles. The first property excludes the existence of $\alpha^2$ dynamos in radiative zones and the second property makes the existence of $\alpha\Omega$ dynamos (with weak differential rotation in mid-latitudes)  unlikely. As we have shown only a dynamo with  $\alpha \propto \cos\theta$    produces the toroidal fields in mid-latitudes. For $\ell>1$, however, the belts are more and more shifted  into the polar region. Such fields   cannot close the loop of field amplification by reproducing the original axisymmetric toroidal field. 


Or, with other words, the  rotation law with the considered  profile ($\Omega\propto \cos^2\theta$) mainly induces toroidal fields at mid-latitudes where for high values of $\ell$ almost no $\alpha$ effect exists. Hence, an  $\alpha\Omega$ dynamo  could only work for very fast rotation. By very fast rotation the instability is suppressed (see Fig. \ref{f1}). 
This topological  problem seems to be the key problem with the magnetic-driven dynamo rather than the  excitation conditions for $\alpha\Omega$ dynamos. It is, however,  not yet clear whether this argumentation also holds with the same power with  other than the used rotation laws and/or  in fully nonlinear simulations. All the presented $\alpha\Omega$ dynamo models working with a weak   solar-type differential rotation are reproducing the assumed sign of the large-scale current helicity ${\vec B}\cdot \curl {\vec B}$. If thus the dynamo produces its own $\alpha$ effect by the magnetic instability then the signs will be consistent.

A basic deficit of the presented theory is the fact that both helicity and $\alpha$ effect have been computed in rigidly rotating stars while for an $\alpha\Omega$ dynamo the rotation must be nonrigid. It might  be the case that growth rates and magnetic patterns of the Tayler instability are strongly modified by even weak differential rotation. The discussion of this new subject, however, is not the scope of the present paper. We have thus considered only  cases with a rather weak  differential rotation (3 \%).

Another deficit is formed by the order-of-magnitude estimation of the eddy diffusivity. Using the characteristic scales of the modes with the highest growth rates the typical velocity is of  order 1 mm/s and the typical diffusivity value is about $10^5$ cm$^2$/s (both for solar values). 
Estimations of the radial mixing produced by the same instability provide an important test of the theory. The observed content of light elements at the Sun impose restrictions on radial mixing in the upper radiative core (Barnes et al.  1999, and references therein). The chemical mixing  in the deep stellar interior can also be compatible with observations only if its characteristic time is longer than the evolutionary time scale \citep{KW94}. Only very slightly supercritical kink-type instability can satisfy this restriction \citep{KR08}.

 All the material  values in the paper are solar values.
 The formulation of the dynamo theory is such that the material parameters of the stellar interior do not appear (see Eq. (\ref{AA})). The conclusions about the dynamo activity for hot stars do thus not depend on the mass of the star.  This is not true, however, for the above mentioned  order-of-magnitude estimations of the characteristic values of  velocity and diffusivity which are running with $\Omega/N$ and $(\Omega/N)^2$, resp. From Fig. \ref{f5} one finds that the given solar values increase by  factors of 30 or 900, resp., for stars with three solar masses.

Only  nonlinear simulations can demonstrate whether the system indeed prefers the modes with the highest growth rates. Only if  not then the  radiative-zone dynamo has  a  chance to work in the solar interior  but if it  works then  it should be only marginally supercritical.

\section*{Acknowledgments}
LLK is grateful to the Deutsche Forschungsgemeinschaft for the  support of the project (436 RUS 113/839). 

\bsp

\label{lastpage}

\end{document}